\documentclass[twocolumn,10pt]{asme2ej}

\usepackage{amsmath,amssymb,amsfonts,amsthm}
\usepackage{gensymb}
\usepackage{balance}
\usepackage{graphicx}
\usepackage{subcaption}
\usepackage{cite}
\usepackage[dvipsnames,usenames]{color}
\usepackage[colorlinks, linkcolor=Blue, filecolor =Red,citecolor=RoyalPurple]{hyperref}
\usepackage{scalerel}
\usepackage{nopageno}

\usepackage{adjustbox}
\usepackage[ruled,vlined]{algorithm2e}
\usepackage{algpseudocode}

\makeatletter
\def\endthebibliography{%
	\def\@noitemerr{\@latex@warning{Empty `thebibliography' environment}}%
	\endlist
}
\makeatother

\usepackage{hyperref}
\hypersetup{
	colorlinks=true,
	linkcolor=blue,
	filecolor=blue,
	citecolor = red,      
	urlcolor=cyan,
}

\usepackage{hyperref}

\newcommand{\markedManu}{unMarked} 
\newcommand{\version}{arxiv} 

\newboolean{showArxiv}
\newboolean{showArxivAlt}

\ifthenelse{\equal{\version}{journal}}
{
	\setboolean{showArxiv}{false} 
	\setboolean{showArxivAlt}{true}
}

\ifthenelse{\equal{\version}{arxiv}}
{
	\setboolean{showArxiv}{true} 
	\setboolean{showArxivAlt}{false}
}

\ifthenelse{\equal{\markedManu}{MARKED}}
{
	\def\rd#1{{\color{red}{#1}}}
	
}

\ifthenelse{\equal{\markedManu}{unMarked}}
{
	\def\rd#1{{\color{black}{#1}}}
	
}

\def\plantFullname{District Cooling Energy Plant}
\def\plant{DCEP}
\def\plants{DCEPs}

\def\triangleq{\ensuremath{\stackrel{\Delta}{=}}}
\def\R{\mathbb{R}}

\def\eqdef{\ensuremath{:=}}

\def\sim{\ensuremath{\mathrm{sim}}}

\def\stateSet{{\sf{X}}}
\def\inputSet{{\sf{U}}}

\def\numPolImp{{\sf{N}}\ensuremath{_{\text{pol}}}}
\def\timeHorzRL{{\sf{T}}\ensuremath{_{\text{sim}}}}
\def\timeHorzMPC{{\sf{T}}\ensuremath{^{\text{plan}}}}
\def\switch{\ensuremath{\mathrm{switch}}}
\def\Z{\ensuremath{\mathbb{Z}}}

\def\ct{\text{ct}}
\def\sw{\text{sw}}
\def\rw{\text{rw}}
\def\chw{\text{chw}}
\def\chws{\text{chw,s}}
\def\chwr{\text{chw,r}}
\def\bp{\text{bp}}
\def\ch{\text{ch}}
\def\cwr{\text{cw,r}}
\def\cw{\text{cw}}
\def\cws{\text{cw,s}}

\def\oawb{\text{oawb}}
\def\oa{\text{oa}}

\def\pw{\text{pw}}
\def\pump{\text{pump}}

\def\lw{\text{lw}}
\def\lws{\text{lw,s}}
\def\lwr{\text{lw,r}}
\def\indv{\text{indv}}
\def\twc{\text{twc}}
\def\tww{\text{tww}}
\def\tw{\text{tw}}
\def\UB{\text{UB}}

\def\set{\text{set}}
\def\TES{\text{TES}}

\def\qDotL{\ensuremath{\dot{q}^\mathrm{L}}}
\def\qDotLref{\ensuremath{\dot{q}^\mathrm{L,ref}}}
\def\qDotCh{\ensuremath{\dot{q}^\mathrm{ch}}}
\def\qDotCond{\ensuremath{\dot{q}^{\text{cond}}}}
\def\qDotCT{\ensuremath{\dot{q}^\mathrm{ct}}}
\def\qDotCTUB{\ensuremath{\dot{q}^{\ct}_{\UB}}}
\def\qDotCTUBk{\ensuremath{\dot{q}^{\ct}_{\UB,k}}}
\def\qChCapIndiv{\ensuremath{\dot{q}^{\ch}_{\indv}}}

\def\mDotCW{\ensuremath{\dot{m}^{\cw}}}
\def\mDotLW{\ensuremath{\dot{m}^{\lw}}}
\def\mDotTW{\ensuremath{\dot{m}^{\tw}}}
\def\mDotSW{\ensuremath{\dot{m}^{\sw}}}
\def\mDotOA{\ensuremath{\dot{m}^{\oa}}}
\def\mDotBP{\ensuremath{\dot{m}^{\bp}}}

\def\nCh{\ensuremath{n^{\text{ch}}}}
\def\nChMax{\ensuremath{n_\text{max}^{\ch}}}

\def\TCWR{\ensuremath{T^{\cwr}}}
\def\TCWS{\ensuremath{T^{\cws}}}
\def\TOAWB{\ensuremath{T^{\text{oawb}}}}

\def\TLWR{\ensuremath{T^{\lwr}}}
\def\TChWS{\ensuremath{T^{\chws}}}

\def\TChWSMin{\ensuremath{T^{\chws}_{\text{min}}}}
\def\TChWSMax{\ensuremath{T^{\chws}_{\text{max}}}}
\def\TCWR{\ensuremath{T^{\cwr}}}
\def\TCWRMax{\ensuremath{T^{\cwr}_{\text{max}}}}
\def\TCWS{\ensuremath{T^{\cws}}}

\def\TLWRMax{\ensuremath{T^{\lwr}_{\text{max}}}}

\def\pb#1{\notes{pb: \rd{#1}      }}

\newlength{\noteWidth}
\long\def\notes#1{\ifinner
	{\footnotesize #1}
	\else
	\marginpar{\parbox[t]{\noteWidth}{\raggedright\footnotesize #1}}
	\fi\typeout{#1}}

\usepackage{xparse}

\ExplSyntaxOn

\ExplSyntaxOff
\ExplSyntaxOn
\cs_new:Nn \pb_prop_gset_bykeys:Nn
{
	\prop_clear:N \l__pb_temp_prop
	\keys_set:nn { pb/propbykey } { #2 }
	\prop_gset_eq:NN #1 \l__pb_temp_prop
}
\cs_generate_variant:Nn \pb_prop_gset_bykeys:Nn { c }

\keys_define:nn { pb/propbykey }
{
	unknown .code:n = \prop_put:Nxn \l__pb_temp_prop { \l_keys_key_tl } { #1 }
}

\cs_generate_variant:Nn \prop_put:Nnn { Nx }

\NewDocumentCommand{\setupcollaborator}{mm}
{
	\prop_new:c { g_collaborator_#1_prop }
	\pb_prop_gset_bykeys:cn { g_collaborator_#1_prop } { #2 }
}

\NewDocumentCommand{\selectcollaborator}{m}
{
	\prop_map_inline:cn { g_collaborator_#1_prop }
	{
		\tl_set:cn { ##1 } { ##2 }
	}
}
\ExplSyntaxOff

\title{Optimal Control of \plantFullname\ with Reinforcement Learning and MPC}

\author{Zhong Guo\thanks{Corresponding author. The research reported here has been partially supported by the NSF through award 1934322 (CMMI) and 2122313 (ECCS).} 
	\affiliation{PhD candidate\\
		Department of Mechanical Engineering\\
		University of Florida\\
		Gainesville, Florida 32601\\
		Email: zhong.guo@ufl.edu
	}	
}

\author{Aditya Chaudhari 
		\affiliation{Postdoctoral Researcher\\
		Department of Mechanical Engineering\\
		University of Florida\\
		Gainesville, Florida 32601\\
		Email:ad.chaudhari@ufl.edu
	}
}
\author{Austin R. Coffman
	\affiliation{PhD\\ 
		Department of Mechanical Engineering\\ 
		University of Florida\\
		Gainesville, Florida 32601\\ 
		Email: bubbaroney@ufl.edu
	}
}

\author{Prabir Barooah
	\affiliation{ Professor\\
		Department of Electronics and Electrical Engineering\\
		Indian Institute of Technology (Guwahati)\\
		Guwahati, Assam, 781039 \\
		India \\
		Email: pbarooah@iitg.ac.in
	}
}

\begin{document}
	
	\maketitle    
	
	\begin{abstract}
		{\it We consider the problem of optimal control of district cooling energy plants (\plants) consisting of multiple chillers, a cooling tower, and a thermal energy storage (TES), in the presence of time-varying electricity price. A straightforward application of model predictive control (MPC) requires solving a challenging mixed-integer nonlinear program (MINLP) because of the on/off of chillers and the complexity of the \plant\ model. Reinforcement learning (RL) is an attractive alternative since its real-time control computation is much simpler. But designing an RL controller is challenging due to myriad design choices and computationally intensive training.
			
			In this paper, we propose an RL controller and an MPC controller for minimizing the electricity cost of a \plant, and compare them via simulations. The two controllers are designed to be comparable in terms of objective and information requirements. The RL controller uses a novel Q-learning algorithm that is based on least-squares policy iteration. We describe the design choices for the RL controller, including the choice of state space and basis functions, that are found to be effective. The proposed MPC controller does not need a mixed integer solver for implementation, but only a nonlinear program (NLP) solver. A rule-based baseline controller is also proposed to aid in comparison. Simulation results show that the proposed RL and MPC controllers achieve similar savings over the baseline controller, about 17\%.
		}
	\end{abstract}
	
	
	
	\section{Introduction}
	
\begin{figure}
		\centering
		\includegraphics[width=1\columnwidth]{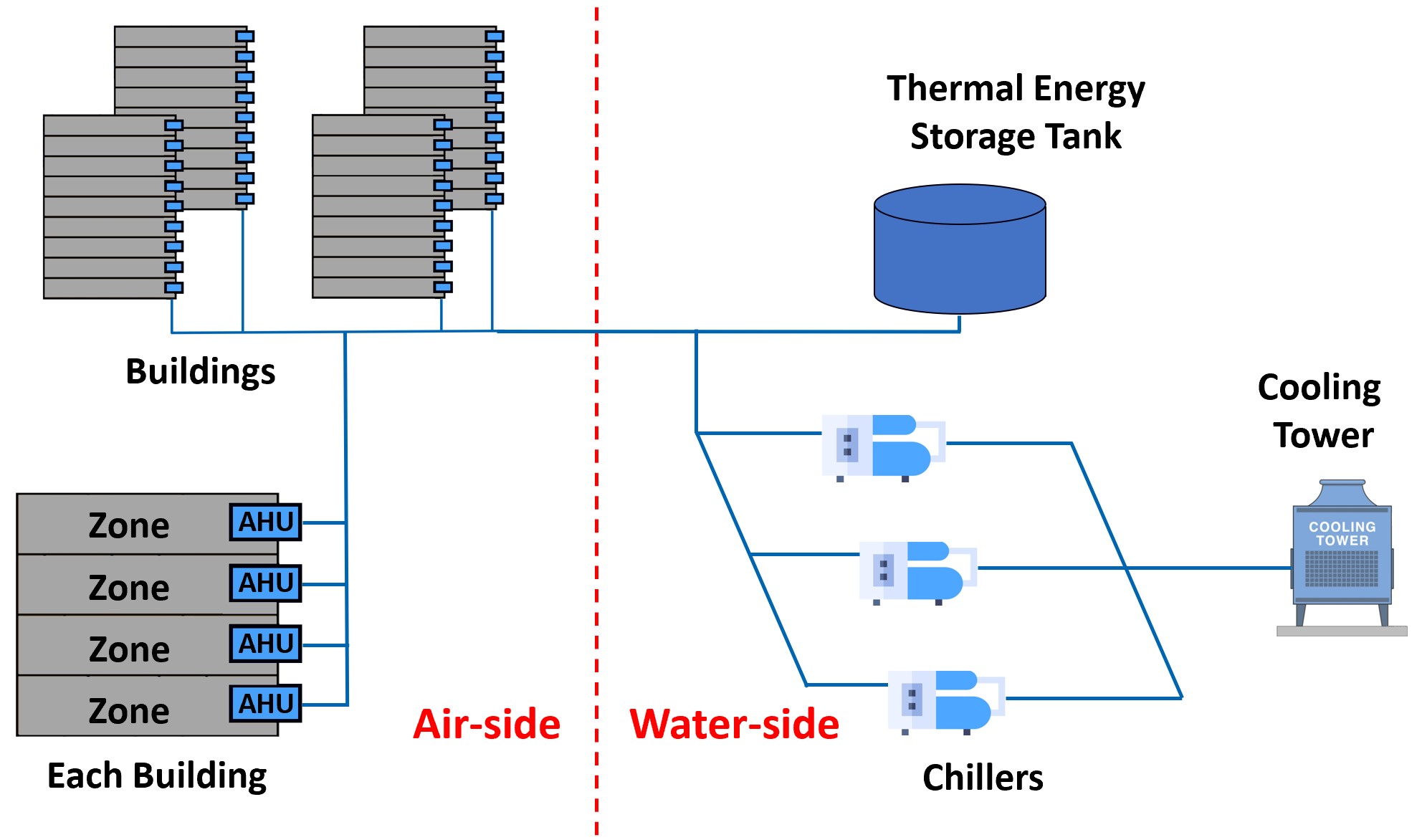}
		\caption{Layout of \plantFullname.}
		\label{fig:DCEP}
	\end{figure}

	In the U.S., 75\% of the electricity is consumed by buildings, and a large part of that is due to heating, ventilation, and air conditioning (HVAC) systems~\cite{CBECS:12}. In university campuses and large hotels, a large portion of the HVAC's share of electricity is consumed by \plantFullname s (\plants), especially in hot and humid climates. A \plant\ produces and supplies chilled water to a group of buildings it serves (hence the moniker ``district''), and the air handling units in those buildings use the chilled water to cool and dehumidify air before supplying it to building interiors. Figure~\ref{fig:DCEP} shows a schematic of such a plant, which consists of multiple chillers that produce chilled water, a cooling tower that rejects the heat extracted from chillers to the environment, and a thermal energy storage system (TES) for storing chilled water. Chillers - the most electricity intensive equipment in the \plant\ -  can produce more chilled water than buildings' needs when electricity price is low. The extra chilled water is then stored in the TES, and used during periods of high electricity price to reduce the total electricity cost. The \plantFullname s are also called central plants or chiller plants. 
	
	\plants\ are traditionally operated with rule-based control algorithms that use heuristics to reduce electricity cost while meeting the load, such as ``chiller priority'', ``storage priority'', and additional control sequencing for the cooling tower operation~\cite{PGNEThermal:1997,HydemanASHRAE:2007,Teleke_TransSustEner_2010,Tam_IntlHPB2018,Pinamonti_Ener2020,Lee_Ener2015,Schibuola_EnerAndBldg2015}. But making the best use of the chillers and the TES to keep the electricity cost at the minimum requires non-trivial decision making due to the discrete nature of some control commands, such as
	chiller on/off actuation, and highly nonlinear dynamics of the equipment in \plants.  A growing body of work has proposed algorithms for optimal real-time control of \plants. Both Model Predictive Control (MPC)~\cite{Ma_CSM:12,ColeMPCACC:2012,TouretzkyMPCJPS:2014,Zabala_RenewAndSustainEnergyReview2020_DCEPNLP,RisbeckMixedintegerEnB:2017,rawlings2018economic,PatelCase:2018,DengMPCASE:2014,KimSiteAE:2022} and Reinforcement Learning (RL)~\cite{Manoharan_ACMConfOnSysForEnergyEffiBldg2021_DCEPRL, QiuModelfreeSTBE:2020, Qiu_EnergyAndBldg2022_DCEPOpti, Nagarathinam_ProcedOf11ACMIntlConfOnFutureEnergySys2020_DCEPRL, Campos_IndustrialAndEngChemResrch2022_DCEPRL, Ahn_SciAndTechForBultEnviron2019_DCEPRL, QiuModelfreeEB:2020,HenzeEvaluationHVACRR:2003, LiuEvaluationJSEE:2007} have been studied. 
	
	For MPC, a direct implementation requires solving a high dimension mixed-integer linear program (MINLP) that is quite challenging to solve. Various substitutive approaches are thus used, which can be categorized into two groups: NLP approximations~\cite{Ma_CSM:12,ColeMPCACC:2012,TouretzkyMPCJPS:2014,Zabala_RenewAndSustainEnergyReview2020_DCEPNLP} and MILP approximations~\cite{RisbeckMixedintegerEnB:2017,rawlings2018economic,DengMPCASE:2014,PatelCase:2018,KimSiteAE:2022}. NLP approximations generally leave the discrete commands for some predetermined control logic and only deal with continuous control commands, which may limit the potential of their savings. MILP approximations mostly adopt a linear \plant\ model so that the problem is tractable, though solving large MILPs is also challenging.
	
	An alternative to MPC is Reinforcement Learning (RL): an umbrella term for a set of tools used to approximate an optimal policy using data	collected from a physical system, or more frequently, its simulation. Despite a burdensome design and learning phase, real-time control is simpler since control computation is an evaluation of a	state-feedback policy. 
	However, designing an RL controller for a \plant\ is quite challenging. The	performance of an RL controller depends on many design choices and training an RL controller is computationally onerous. 
	
	In this paper we propose an RL controller and an MPC controller for a \plant, and compare their performance with that of a rule-based baseline (BL) controller through simulations. All three controllers are designed to minimize total energy cost while meeting the required cooling load. The main source of flexibility is the TES, which allows a well-design controller to charge the TES in periods of low electricity price. The proposed RL controller is based on a new learning algorithm that is inspired by the ``convex Q-learning'' proposed in recent work~\cite{ConvexLuACC:2021} and the classical least squares policy iteration (LSPI) algorithm~\cite{LeastSquaresLagoudakisJMLR:2003}. Basis functions are carefully designed to reduce computational burden in training the RL controller. The proposed MPC controller solves a two-fold non-linear program (NLP) that is transformed from the original MINLP via heuristics. Hence the MPC controller is ``stand-in'' for a true optimal controller and provides a sub-optimal solution to the original MINLP. The baseline controller that is used for comparison is designed to utilize the TES and time varying electricity prices (to the extent possible with heuristics) to reduce energy costs. The RL controller and baseline controller have the same information about electricity price: the current price and a backward moving average.
	
	The objective behind this work is to compare the performance of the two complementary approaches, MPC and RL, for the optimal control of all the principal actuators in a \plant. The two controllers are designed to be comparable, in terms of objective and information requirements. We are not aware of many works that has performed such a comparison; the only exceptions are~\cite{LiuEvaluationJSEE:2007,HenzeEvaluationHVACRR:2003}, but the decision making is limited to a TES or temperature setpoints. Since both RL and MPC approaches have merits and weaknesses, designing a controller with one approach and showing it performs well leaves open the question: would the other have performed better?  This paper takes a first step in addressing such questions. To aid in this comparison, both the controllers are designed to be approximations of the same intractable infinite horizon optimal control problem. Due to the large difference in the respective approaches (MPC and RL), it is not possible to ensure exact parallels for an``apples--to--apples'' comparison. But the design problems for RL and MPC controllers have been formulated to be similar to the possible extent.
	
	Simulations results show that both the controllers, RL and MPC, leads to significant and similar cost savings (16-18\%) over a rule-based baseline controller. These values are comparable to that of MPC controllers with mixed-integer formulation reported in the	literature, which vary from 10\% to 17\%~\cite{RisbeckMixedintegerEnB:2017,rawlings2018economic,DengMPCASE:2014,PatelCase:2018,KimSiteAE:2022}. The cooling load tracking performance is similar between them. The real-time computation burden of the RL controller is trivial compared to that of the MPC controller, but the RL controller leads to higher chiller switches (from off to on and vice versa). However, the MPC controller enjoys the advantage of error-free forecasts in the simulations, something the RL controller does not.
	
	The rest of the manuscript is organized as follows. The contribution of the paper over the related literature is discussed in detail in Section~\ref{sec:lit}. Section~\ref{sec:sysDesc} describes the \plantFullname\ and its simulation model as well as the control problem. Section~\ref{sec:RL} describes the proposed RL controller, Section~\ref{sec:MPC} the proposed MPC controller, and Section~\ref{sec:ruleBased} describes the baseline controller. Section~\ref{sec:eval} provides simulation evaluation of the controllers. Section~\ref{sec:under-the-hood} provides an ``under-the-hood'' view of the design choices for the RL controller. Section~\ref{sec:conclusion} concludes the paper.
	
	\subsection{Literature Review and Contributions}\label{sec:lit}
	\subsubsection{Prior work on RL for \plant} 
	There is a large and growing body of work in this area, e.g.~\cite{HenzeEvaluationHVACRR:2003,Manoharan_ACMConfOnSysForEnergyEffiBldg2021_DCEPRL, QiuModelfreeSTBE:2020, Qiu_EnergyAndBldg2022_DCEPOpti, Nagarathinam_ProcedOf11ACMIntlConfOnFutureEnergySys2020_DCEPRL, Campos_IndustrialAndEngChemResrch2022_DCEPRL, Ahn_SciAndTechForBultEnviron2019_DCEPRL, QiuModelfreeEB:2020, LiuEvaluationJSEE:2007}. Most of these papers limit the problem to controlling part of a \plant. For instance, the \plant s considered in~\cite{Manoharan_ACMConfOnSysForEnergyEffiBldg2021_DCEPRL, QiuModelfreeSTBE:2020,Qiu_EnergyAndBldg2022_DCEPOpti,Nagarathinam_ProcedOf11ACMIntlConfOnFutureEnergySys2020_DCEPRL,Ahn_SciAndTechForBultEnviron2019_DCEPRL} do not have a TES. 
	Refs.~\cite{Manoharan_ACMConfOnSysForEnergyEffiBldg2021_DCEPRL,QiuModelfreeSTBE:2020,Qiu_EnergyAndBldg2022_DCEPOpti,Campos_IndustrialAndEngChemResrch2022_DCEPRL,Nagarathinam_ProcedOf11ACMIntlConfOnFutureEnergySys2020_DCEPRL} optimize only the chilled water loop but not the cooling water loop (at the cooling tower), while~\cite{QiuModelfreeEB:2020} only optimize the cooling water loop. The reported energy savings are in the 10-20\% range over rule-based baseline controllers; e.g. 15.7\% in~\cite{Ahn_SciAndTechForBultEnviron2019_DCEPRL}, 11.5\% in~\cite{Manoharan_ACMConfOnSysForEnergyEffiBldg2021_DCEPRL} and around 17\% in~\cite{Nagarathinam_ProcedOf11ACMIntlConfOnFutureEnergySys2020_DCEPRL}.

	The ref.~\cite{HenzeEvaluationHVACRR:2003} considers a complete \plant, but the control command computed by the RL agent is limited to TES charging and discharging. It is not clear what control law is used to decide chiller commands and cooling water loop setpoints. The work~\cite{LiuEvaluationJSEE:2007} also considers a complete \plant, with two chillers, a TES, and a large building with an air handling unit. The RL controller is tasked with commanding only the zone temperature setpoint and TES charging/discharging flowrate whilst the control of the chillers or the cooling tower is not considered. Besides, trajectories of external inputs, e.g., outside air temperature and electricity price, are the same for all training days in~\cite{LiuEvaluationJSEE:2007}. Another similarity of~\cite{LiuEvaluationJSEE:2007,HenzeEvaluationHVACRR:2003} with this paper is that these references compare the performance of RL with that of a model-based predictive control.
	
	\subsubsection{Prior work on MPC for \plant} 
	The works that are closest to us in terms of problem setting are~\cite{RisbeckMixedintegerEnB:2017,PatelCase:2018,rawlings2018economic}, which all reported MILP relaxation-based MPC schemes to optimally operate a \plant\ with TES in presence of time varying electricity prices. The paper~\cite{RisbeckMixedintegerEnB:2017} reports an energy cost savings with MPC of about 10\% over a baseline strategy that uses a common heuristic (charge TES all night) with some decisions made by optimization. In~\cite{rawlings2018economic}, around 15\% savings over the currently installed rule-based controller is achieved in a real \plant. The study \cite{PatelCase:2018} reported a cost savings of by 17\% over ``without load shifting'' with help of the TES in a week-long simulation. The paper~\cite{DengMPCASE:2014} also proposes an MILP relaxation based MPC scheme for controlling a \plant\ and approximately 10\% savings in electricity cost over a baseline controller over a one-day long simulation is reported. But the \plant\ model in~\cite{DengMPCASE:2014} ignores the effect of weather condition on plant efficiency, and the baseline controller is not purely-rule based; it makes TES and chiller decisions based on a greedy search. The recent paper~\cite{KimSiteAE:2022} deserves special mention since it reports an experimental demonstration of MPC applied to a large \plant; the control objective being manipulation of demand to help with renewable integration and decarbonization. It too uses an MILP relaxation. The decision variables include plant mode (combination of chillers on) and TES operation, but cooling water loop decisions are left to legacy rule-based controllers. 
	
	There is another body of work applying MPC to the control a \plant, such as~\cite{Ma_CSM:12,Zabala_RenewAndSustainEnergyReview2020_DCEPNLP,ColeMPCACC:2012,TouretzkyMPCJPS:2014}. But they either ignore the on/off nature of the chiller control~\cite{Ma_CSM:12,ColeMPCACC:2012} or reformulate the problem using some heuristics~\cite{Zabala_RenewAndSustainEnergyReview2020_DCEPNLP,TouretzkyMPCJPS:2014} so that the underlying optimization problem is naturally an NLP.
	
	\subsection{Contribution over Priori Arts}
	\subsubsection{Contribution over ``RL for \plant'' literature:}
	Unlike most prior works on RL for \plants\ that only deal with a part of \plant\ \cite{Manoharan_ACMConfOnSysForEnergyEffiBldg2021_DCEPRL, QiuModelfreeSTBE:2020, Qiu_EnergyAndBldg2022_DCEPOpti, Nagarathinam_ProcedOf11ACMIntlConfOnFutureEnergySys2020_DCEPRL, Campos_IndustrialAndEngChemResrch2022_DCEPRL, Ahn_SciAndTechForBultEnviron2019_DCEPRL, QiuModelfreeEB:2020}, the control commands in this work consist of all the available commands (five in total) of both the water loops in a full \plant. To the best of our knowledge no prior work has used RL to command both the water loops and a TES. Second, unlike some of the closely related work such as \cite{LiuEvaluationJSEE:2007}, we treat external inputs such as weather and electricity price as RL states, making the proposed RL controller applicable for any time-varying disturbances that can be measured in real time. Otherwise the controller is likely to work well only for disturbances seen during training. Third, the proposed RL controller commands the on/off status of chillers directly rather than the chilled/cooling water temperature setpoints~\cite{QiuModelfreeSTBE:2020,Ahn_SciAndTechForBultEnviron2019_DCEPRL,Nagarathinam_ProcedOf11ACMIntlConfOnFutureEnergySys2020_DCEPRL} or zone temperature setpoints~\cite{LiuEvaluationJSEE:2007}, which eliminates the need for another control system to translate those setpoints to chiller commands. Fourth, all the works cited above rely on discretizing the state and/or action spaces in order to use the classical tabular learning algorithms with the exception of \cite{Campos_IndustrialAndEngChemResrch2022_DCEPRL}. The size of the table will become prohibitively large if the number of states and control commands become large and a fine resolution discretization is used. Training a such controller and using it in real time, which will require searching over this table, will become computationally challenging. That is perhaps why only a small number of inputs are chosen as control commands in prior work even though several more setpoints can be manipulated in a real \plant. Although~\cite{Campos_IndustrialAndEngChemResrch2022_DCEPRL} considers continuous states, its proposed method only controls part of a \plant\ with simplified linear plant model, which may significantly limit its potential of cost savings in reality. In contrast, the RL controller proposed in this paper is for a \plant\ model consisting of highly nonlinear equations and the states and actions are kept as continuous except for the one command that is naturally discrete (number of chillers that are on). 
	
	While there is an extensive literature on learning algorithms and on designing RL controllers, design of an RL controller for practically relevant applications with non-trivial dynamics is quite challenging. RL's performance depends on myriad design choices, not only on the stage cost/reward, function approximation architecture and basis functions, learning algorithm and method of exploration, but also on the choice of the state space itself. A second challenge is that training a RL controller is computationally intensive and brute force training is beyond the computational means of most researchers. For instance, The hardware cost for a single AlphaGo Zero system in 2017 by DeepMind has been quoted to be around \$25 million~\cite{GibneySelfNature:2017}. Careful selection of the design choices mentioned above is thus required, which leads to the third challenge: if a particular set of design choices lead to a policy that does not perform well, there is no principled method to look for improvement. Although RL is being extensively studied in the control community, most works demonstrate their algorithms on plants with simple dynamics with a small number of states and inputs; e.g.~\cite{DataDrivenBanjacCDC:2019,PolicyLuoToC:2017}. The model for a \plant\ used in this paper, arguably still simple compared to available simulation models (e.g.~\cite{FanOpensourceAE:2021}), is quite complex: it has 8 states, 5 control inputs, 3 disturbance inputs, and requires solving an optimization problem to compute the next state given the current state, control and disturbance.
	
	\subsubsection{Contribution over ``MPC for \plant'' literature:} The MPC controller proposed here uses a combination of relaxation and heuristics to avoid the MINLP formulation. In contrast to~\cite{DengMPCASE:2014,PatelCase:2018,rawlings2018economic,RisbeckMixedintegerEnB:2017,KimSiteAE:2022}, the MPC controller does not use a MILP relaxation. The controller does compute discrete decisions (number chillers to be on, TES charge/discharge) directly, but it does so by using NLP solvers in conjunction with heuristics. The cost savings obtained is similar to those reported in earlier work that use MILP relaxation. Comparing other NLP formulations~\cite{Ma_CSM:12,Zabala_RenewAndSustainEnergyReview2020_DCEPNLP,ColeMPCACC:2012,TouretzkyMPCJPS:2014}, our MPC controller determines the on/off actuation of equipments and TES charging/discharging operation directly. 
	
	Closed loop simulations are provided for all three controllers, RL, MPC, and baseline, to assess the trade-offs among these controllers, and especially between the model-based MPC controller and the ``model-free'' RL controller. 
	
	\subsubsection{Contribution over a preliminary version:} The RL controller described here was presented in a preliminary version of this paper~\cite{GuoReinforcementACC:2022}. There are three improvements. Firstly, a MPC controller, which is not presented in~\cite{GuoReinforcementACC:2022}, was designed, evaluated, and compared with our RL controller. Therefore, the optimality of our control with RL is better assessed. Another difference is that the baseline controller described here is improved over that in~\cite{GuoReinforcementACC:2022} so that frequency on/off switching of chillers is reduced. Lastly, a much more thorough discussion of the RL controller design choices and their observed impacts are included here than in~\cite{GuoReinforcementACC:2022}. Given the main challenge with designing RL controllers for complex physical systems discussed above, namely, ``what knob to tweak when it doesn't work?'', we believe this information will be valuable to other researchers. 
	\section{System description and control problem}\label{sec:sysDesc}
\begin{figure*}[h]
		\centering
		\includegraphics[width=1.725\columnwidth, height=0.82\columnwidth]{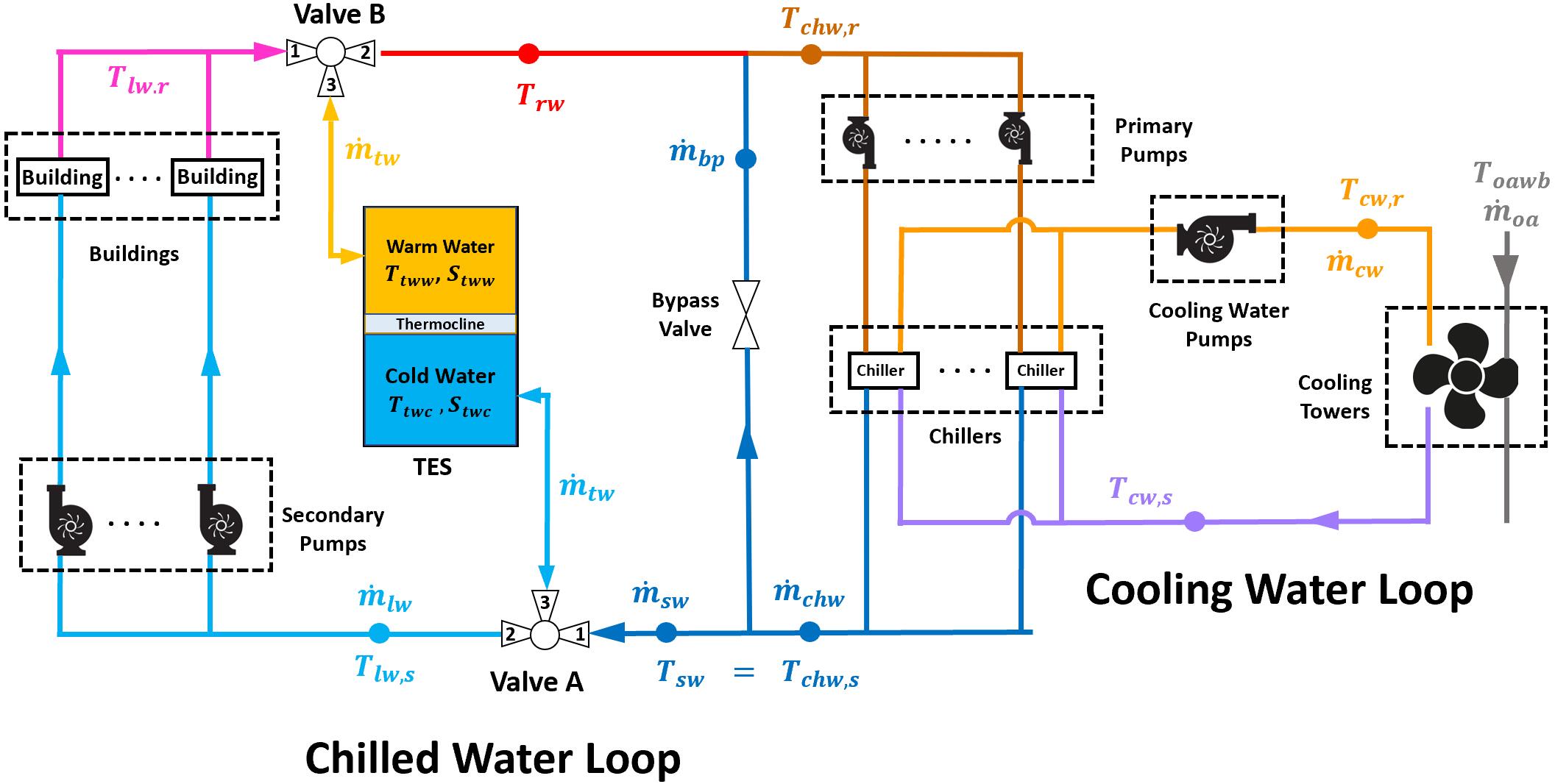}
		\caption{Detailed description of \plantFullname.}
		\label{fig:ChillerPlant}
	\end{figure*}

	The \plant\ contains a TES, multiple chillers and chilled water pumps, a cooling tower and cooling water pumps, and finally a collection of buildings that uses the chilled water to provide air conditioning; see Figure~\ref{fig:ChillerPlant}. 
	The heat load from the buildings is absorbed by the cold chilled water supplied by the \plant, and thus the return chilled water temperature is warmer. This part of the water is called \emph{load water}, and the related variables are denoted by superscript $\lw$ for ``load water''. The \emph{chiller loop} (subscript $\ch$) removes this heat and transmits it to the \emph{cooling water loop} (subscript $\cw$). The cooling water loop absorbs this heat and sends it to the cooling tower, where this heat is then rejected to the ambient. The cooling tower cool downs the cooling water returned from the chiller by passing both the sprayed cooling water and ambient air through a fill. During this process, a small amount of water spray will evaporate into the air, removing heat from the cooling water. The cooling water loss due to its evaporation is replenished by fresh water, thus we assume the supply water flow rate equals to the return water flow rate at the cooling tower. A fan or a set of fans is used to maintain the ambient airflow at the cooling tower.  Connected to the chilled water loop is a TES tank that stores water (subscript $\tw)$. The total volume of the water in the TES tank is constant, but a thermocline separates two volumes: cold water that is supplied by the chiller (subscript $\twc$ for ``tank water, cold'') and warm water returned from the load (subscript $\tww$ for ``tank water, warm'').
	
	\subsection{\plant\ dynamics}\label{sec:plantModel}
	Time is discretized with sampling period $t_s$ with a counter $k = 0,1,...$ denoting the time step. With the consideration of hardware limits and ease of implementation, the control commands are chosen as follows: 
	\begin{enumerate}
		\item $\mDotLW_k$, the chilled water flowrate going through the cooling coil, to ensure the required cooling load is met.
		\item  $\mDotTW$, charging/discharging flowrate of the TES, to take the advantage of load shifting.
		\item $\nCh$, the number of active chillers to ensure the amount of chilled water required is met and the coldness of the chilled water is maintained.
		\item $\mDotCW$, the flowrate of cooling water going through the condenser of chillers to absorb the heat from the chilled water loop. 
		\item $\mDotOA$, the flowrate of ambient air that cools down the cooling water to maintain its temperature within the desired range.
	\end{enumerate} 
	Therefore, the control command $u_k$ is:
	\begin{align}
		u_k := [\mDotLW_k, \mDotTW_k, \nCh_k, \mDotCW_k, \mDotOA_k]^T \in\inputSet.
	\end{align}
	Each of these variables can be independently chosen as setpoints since lower level PI-control loops maintain them. There are limits to these setpoints, which determine the admissible input set $\inputSet \subset \{0,\dots, \nChMax\} \times \R^4$:
	\begin{align} \nonumber
		\inputSet &\triangleq  \{0,\dots, \nChMax\} \times [\dot{m}_\text{min}^{\lw},\dot{m}_\text{max}^{\lw}] \times [\dot{m}_\text{min}^{\tw},\dot{m}_\text{max}^{\tw}] \dots \\ \times &[\dot{m}_\text{min}^{\cw},\dot{m}_\text{max}^{\cw}] \times [\dot{m}_\text{min}^{\oa},\dot{m}_\text{max}^{\oa}]\subset \R^5.
	\end{align}
	Since the TES can be charged and discharged, we declare $\mDotTW>0$ for charging and $\mDotTW<0$ for discharging as a convention. 
	
	The state of the \plant\ $x^p$ is:
	\begin{align}
		\label{eq:stateVecPlant}
		x^p_k \triangleq[T_k^{\lwr}, S_k^\tww, S_k^\twc, T_k^\twc, T_k^\tww,T_k^{\chws}, T_k^{\cwr}, T_k^{\cws}]^T,
	\end{align}
	where $S^\tww, S^\twc$ are the fractions of the warm water and cold water in the TES tank, $S^\tww + S^\twc=1$. The other state variables are temperatures at various locations - supply (subscript ``$,s$'') and return (subscript ``$,r$") - in each water loop: load water, cooling water, tank water, and chiller; see Figure~\ref{fig:ChillerPlant} for details. All the plant state variables $x^p$ can be measured with sensors. The superscript ``$p$" of $x$ emphasizes that $x^p$ is the state of the ``plant'', not the state in the reinforcement learning method that will be introduced in Section~\ref{sec:RLarch}.
	
	The plant state $x^p$ is affected by exogenous disturbances $w^p_k:=[\TOAWB_k,\qDotLref_k]^T \in \R^2$, where $\qDotLref_k$ is the required cooling load, the rate at which heat needs to be removed from buildings, and $T^{\oawb}_k$ is the ambient wet bulb temperature. The disturbance $w^p_k$ cannot be ignored, e.g., ambient wet-bulb temperature plays a critical role in cooling tower dynamics. 
	
	The control command and disturbances affect the state through a highly nonlinear dynamic model:
	\begin{align}\label{eq:plant-model-f}
		x^p_{k+1} & = f(x^p_k,u_k,w^p_k),
	\end{align}
	that is described in details in the Appendix. The dynamics~\eqref{eq:plant-model-f} are implicit: there is no explicit function $f(\cdot)$ that can be evaluated to obtain $x_{k+1}$. The reason is that all the heat exchangers (in each chiller, in the cooling tower, and in the cooling coils in the buildings) has a limited capacity. Depending on the cooling load and the outdoor air wet-bulb temperature, one of the heat exchangers might saturate. Meaning, it will then only deliver as much exchange as its capacity, less than what is desired due to the load. Which heat exchanger will saturate first depends on the current state and disturbance and control on a complex manner.  Hence, some form of iterative computation is required to simulate the dynamics, e.g.,  the method developed in~\cite{YuOptimizationAE:2008}. A generalized way to perform the iterative update to account for the limits of heat exchange capacities is by solving a constrained optimization problem, which is the method used in this work.
	
	The method is described in detail in the Appendix, but here we provide an outline for use in the sequel. First, define the decision variable $z_k$ as
	\begin{align}
		z_k \triangleq \big[(x_{k+1}^p)^T, \ \qDotL_k,\ \qDotCh_k,\ \qDotCT_k\big]^T,
	\end{align}
	where $x^p$ is defined in~\eqref{eq:stateVecPlant}, $\qDotL$ is the cooling load met by the \plant, $\qDotCh$ and $\qDotCT$ are the cooling provided by chillers and cooling towers. The set $\Omega(x_k^p,w^p_k,u_k)$ is defined by the dynamics and
	constraints of the \plant\ system:
	\begin{align} \nonumber
		\Omega(x_k^p,w_k^p, u_k) \triangleq\big\{\mathbf{Z} : \mathbf{Z}
		\text{ satisfies } \eqref{eq:T_lws}-\eqref{eq:CTHeatBalance}\big\}.
	\end{align}
	where the numbered equations inside the definition describe the dynamics of the various heat exchangers and the TES, and the capacity limits of the heat exchangers in the buildings' air handling units, chillers and the cooling tower. The derivation of the equations~\eqref{eq:T_lws}-\eqref{eq:CTHeatBalance} are described in the Appendix. The value of $z_k$ is computed by solving the following optimization problem:
	\begin{align} \label{eq:plantDyn-mainbody}
		z^{*}_k = \arg&\min_{z_{k}\in\Omega(x^p_k,w^p_k,u_k)} r_1\|\qDotL_k-\qDotLref_k\| \nonumber\\
		&+ r_2\|\TChWS_{k+1}-\TChWS_{\set}\| + r_3\|\TCWS_{k+1}-\TCWS_{\set}\|,
	\end{align}
	where $\TChWS_{\set}$ and $\TCWS_{\set}$ are pre-specified setpoints that reflect nominal working conditions and $r_1$, $r_2$, and $r_3$ are positive design choices, with $r_1 \gg r_2, r_3$. When the required cooling load $\qDotLref_k$ is within the capacity of all the heat exchangers, then the solution to~\eqref{eq:plantDyn-mainbody} yields $\qDotL_k = \qDotLref_k$. When the required load exceeds capacity of the \plant, then~\eqref{eq:plantDyn-mainbody} will lead to a solution that trades off maintaining nominal setpoints and meeting the cooling load, while respecting the limits of the heat exchangers. The solution leads to the next state $x^p_{k+1}$ (as the first component of $z_k^*$), ad thus~\eqref{eq:plantDyn-mainbody} implicitly defines the model $f(\cdot)$. In this paper, we use CasADi/IPOPT~\cite{Andersson2019,wacbie:2006} to solve~\eqref{eq:plantDyn-mainbody} for simulating the plant.
	
	\subsection{Electrical demand and electricity cost} \label{sec:Pelec-formulas}
	In the \plant\ considered, the only energy used is electricity. The relationship between the thermal quantities and the electricity consumption in chillers and cooling tower are complex. We model the chillers power consumption $P^{\ch}$ as~\cite{ASHRAE:guideline14_2002}: 
	\begin{align}\label{eq:P_CH}
		P^{\ch}_k & = (\frac{T^{\cws}_k}{T^{\chws}_k}-1)\qDotCh_k-\beta_1+\beta_2T_k^{\cws}-\beta_3\frac{T_k^{\cws}}{T_k^{\chws}}.
	\end{align}
	Power consumption of water pumps is modeled using the black-box model in~\cite{RisbeckMixedintegerEnB:2017}:
	\begin{align}
		P^{\chw,\pump}_k &= \alpha_1\ln(1+\alpha_2\dot{m}_k^{\chw})+\alpha_3\dot{m}_k^{\chw}+\alpha_4, \label{eq:P_chwPumps} 
		\\P_k^{\cw,\pump} &= \gamma_1 \ln(1+\gamma_2\dot{m}_k^{\cw})+\gamma_3\dot{m}^{\cw}_k+\gamma_4.\label{eq:P_cwPumps} 
	\end{align}
	Finally, the electrical power consumption of the cooling tower mainly comes from its fan and is modeled as~\cite{Braun:ASHRAE_Trans_1996}:
	\begin{align} \label{eq:P_CT_mDot_oa}
		P^{\ct}_k = \lambda(\dot{m}^{\oa}_k)^3.
	\end{align}
	The constants $\alpha_i,\beta_i,\gamma_i$, and $\lambda$ are empirical parameters. The total electric power consumption of the \plant\ is:
	\begin{align}
		P^{\text{tot}}_k= P^\ch_k + P^\ct_k + P^{\chw, \pump}_k + P^{\cw, \pump}_k. \label{eq:P_tot}
	\end{align}
	
	\subsection{Model calibration and validation} \label{sec:modelCal}
\begin{figure}[h]
		\centering
		\subcaptionbox{Schematic of the UWC Tampines Campus, from~\cite{MillerSingporeDatabaseUrl:2014}.}
		{\includegraphics[width=0.9\linewidth]{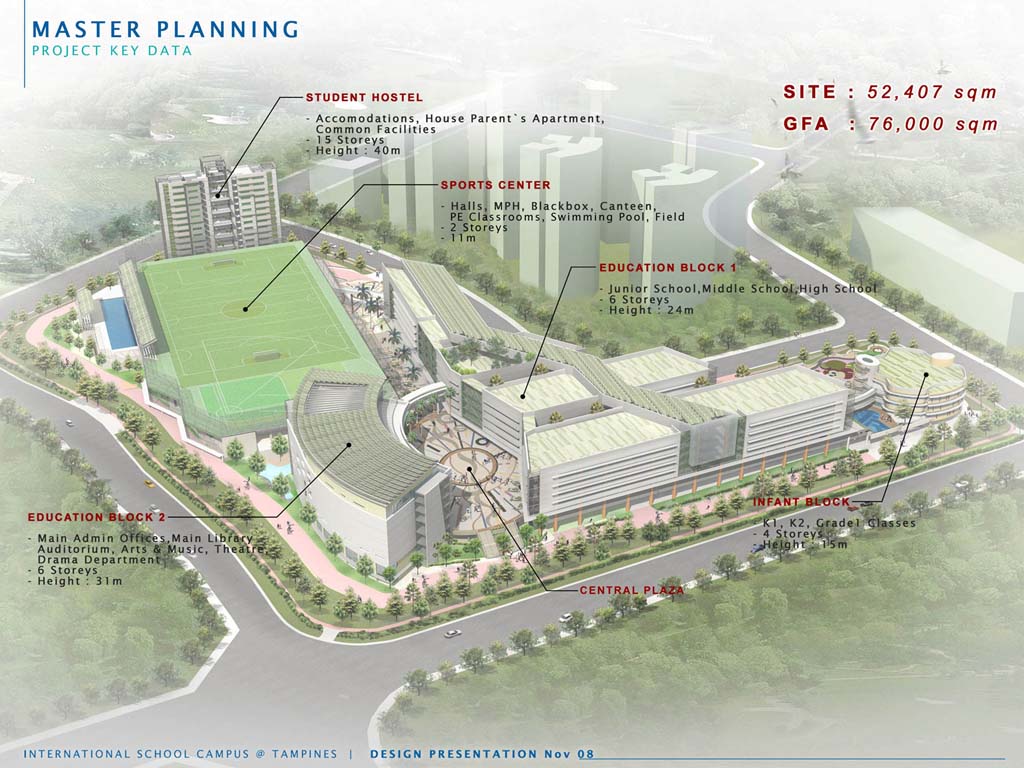}}
		\subcaptionbox{Chiller power measurement $P^{\ch}$ and prediction $\hat{P}^{\ch}$. \label{fig:UWC-campus}}
		{\includegraphics[width=1\linewidth]{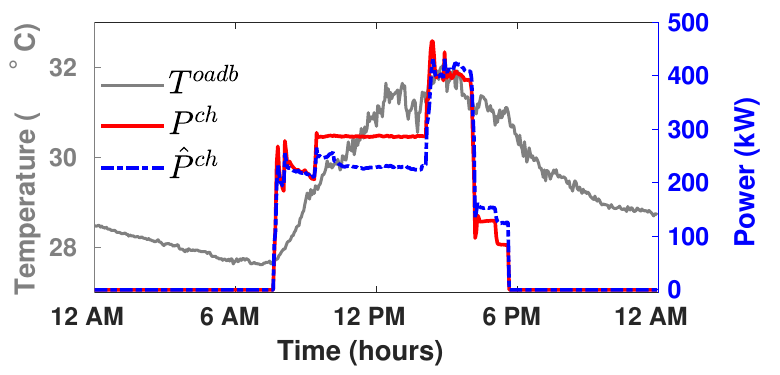}\label{fig:P-ch-calib}}
		\caption{(Top) Map of the campus with a DCEP whose data is
			used for model calibration, and (Bottom) Out of sample prediction for
			$P^{\ch}$ using the calibrated model~\eqref{eq:P_CH}.}\label{fig:Iden_P_CH}
	\end{figure}

	The parameters of the simulation model in Section~\ref{sec:plantModel} and electrical demand model in Section~\ref{sec:Pelec-formulas} are calibrated using data from the energy management system in United World College (UWC) of South East Asia Tampines Campus in Singapore, shown in Figure~\ref{fig:UWC-campus}. The data is publicly available in~\cite{MillerSingporeDatabaseUrl:2014}, and details of the data are discussed in \cite{MillerSingporeDatabasePaper:2014}. There are three chillers and nine cooling towers in the DCEP. The data from chiller one and cooling tower one are used for model calibration. We use 80\% of data for model identification and 20\% of data for verification. The out-of-sample prediction results for the total electrical demand are shown in Figure~\ref{fig:Iden_P_CH}. Comparison between data and prediction for other variables are not shown in the interest of space. 
	
	\subsection{The (ideal) control problem} \label{sec:contGoal} 
	The electricity cost incurred during the $k$-th time step is:
	\begin{align}\label{eq:ckE}
		c^\text{E}_k \eqdef  t_s\rho_k P^{\text{tot}}_k,
	\end{align}
	where $P^{\text{tot}}_k$ is the total electric power consumed in $k$ and is defined in~\eqref{eq:P_tot}. The goal of operating the \plant\ to minimize electricity cost while meeting the required cooling load $\qDotLref_k$ can be posed as the following infinite horizon optimal control problem.
	\begin{align} \label{eq:ControlGoal_infinite} 
		\min_{\{u_k\}_{k=0}^{\infty}} \ &\sum_{k=0}^{\infty}c_k^{\text{E}}, \\
		\text{s.t.} \quad &x^p_{k+1} = f(x^p_k,u_k,w^p_k), \ x^p_0 = x, \nonumber\\
		& x^p_k \in \stateSet^p(w^p_k), \quad u_k \in \inputSet(x^p_k,w_k) \label{eq:SysStateConstraint} \\
		& \qDotL_k(x^p_k,u_k) = \qDotLref_k, \label{eq:MeetLoad} 
	\end{align}
	where $\rho_k$ $(\frac{\text{USD}}{\text{kWh}})$ is the electricity price. The state $x^p_k$, input $u_k$, and disturbance $w^p_k$ of the \plant\ are defined in Section~\ref{sec:plantModel}; $\qDotL_k$ (``L" stands for ``load") represents the actual cooling load met by the \plant, which is a function of $x^p_k$ and $u_k$. The bounds for $x^p_k$ and $u_k$ are $\stateSet^p(w_k)$ and $\inputSet(x^p_k,w_k)$. The reason these sets are dependent on the state or disturbance can be found in the description of the dynamic model of the plant in the Appendix.
	
	Even when truncated to a finite planning horizon considered, Problem~\eqref{eq:ControlGoal_infinite} is an MINLP due to $n^{\ch}_k$ being an integer and the nonlinear dynamics~\eqref{eq:plant-model-f}. In the sequel we propose two controllers to solve approximations of this idealized problem.
	
	\section{RL basics and proposed RL controller}\label{sec:RL}
	
	\subsection{RL basics} \label{sec:RLframe}
	For the following construction, let $x$ represent the state with state space $\stateSet$ and $u$ the input with input space $\inputSet(x)$. Now consider the following infinite horizon discounted optimal control problem
	\begin{align}
		J^*(\bar{x}) = &\min_{\mathbf{U}} \quad \sum_{k=0}^{\infty}\gamma^k c(x_k,u_k), \quad x_0 = \bar{x}, \label{eq:OC_cost} \\
		&\text{s.t.} \quad  x_{k+1} = F(x_k,u_k), \ u_k \in \inputSet(x_k), \nonumber
	\end{align}
	where $\mathbf{U}\triangleq\{u_0,\dots, \}$, $c: \stateSet \times \inputSet \to \R^{\geq 0}$ is the stage cost, $\gamma \in (0,1)$ is the discount factor, $F(\cdot,\cdot)$ defines the dynamics, and $J^* : \stateSet \rightarrow \mathbb{R}^+$ is the optimal value function. The goal of the RL framework is to learn an approximate optimal policy $\phi : \stateSet \rightarrow \inputSet$ for the problem~\eqref{eq:OC_cost} without requiring explicit knowledge of the model $F(\cdot,\cdot)$. The learning process is based on the $Q$ function. Given a policy $\phi$ for the problem~\eqref{eq:OC_cost}, the $Q$ function associated with this policy is defined as
	\begin{align} \label{eq:QfuncFixPol}
		Q_\phi(x,u) = \sum_{k=0}^{\infty}\gamma^k c(x_k,u_k), \quad x_0 = x, \quad u_0 = u,
	\end{align}
	where for $k\geq 0$ we have $x_{k+1} = F(x_k,u_k)$ and for $k \geq 1$ we have $u_k = \phi(x_k)$. 
	A well known fact is that the optimal policy satisfies~\cite{sutbar18}:
	\begin{align} \label{eq:argMinQ}
		\phi^*(x) = \arg\min_{u\in\inputSet(x)} \ Q^*(x,u), \quad \text{for all} \quad x\in\stateSet,
	\end{align}
	where $Q^* \triangleq Q_{\phi^*}$ is the $Q$ function for the optimal policy. Further, for any policy $\phi$ the $Q$ function satisfies the following fixed point relation:
	\begin{align} \label{eq:fixPolBellEq}
		Q_\phi(x,u) = c(x,u) + \gamma Q_\phi\big(x^+, \phi(x^+)\big),
	\end{align}
	for all $u\in\inputSet(x)$, $x\in\stateSet$, and $x^+ = F(x,u)$. The above relation is termed here as the fixed-policy Bellman equation. If the optimal Q-function can be learned, the optimal control command $u^*_k$ is computed from the Q-function as:
	\begin{align} 
		u^*_k := \phi^*(x_k) = \arg\min_{u\in\inputSet(x_k)} \ Q^{*}(x_k,u),
	\end{align}
	
	\subsection{Proposed RL algorithm} \label{sec:RLalg}
	The proposed learning algorithm has two parts: policy evaluation and policy improvement. First, in policy evaluation, a parametric approximation to the fixed policy Q-function is learned by constructing a residual term from~\eqref{eq:fixPolBellEq} as an error to minimize. Second, in policy improvement, the learned approximation is used to define a new policy based on~\eqref{eq:argMinQ}. For policy evaluation, suppose for a policy $\phi$ the $Q$ function is approximated as:
	\begin{align} \label{eq:qHatTheory}
		Q_\phi(x,u) \approx Q_\phi^\theta(x,u)
	\end{align}
	where $Q_\phi^\theta(\cdot,\cdot)$ is the function approximator (e.g., a neural network) and $\theta \in \mathbb{R}^d$ is the parameter vector (e.g., weights of the network). To fit the approximator, suppose that the system is simulated for $\timeHorzRL$ time so that $\timeHorzRL$ tuples of $(x_k,u_k,x_{k+1})$ are collected to produce $\timeHorzRL$ values of:
	\begin{align}
		d_k(\theta) = c(x_k,u_k) + \gamma Q_\phi^\theta(x_{k+1},\phi(x_{k+1})) - Q_\phi^\theta(x_k,u_k),
	\end{align} 
	which is the temporal difference error for the approximator.
	We then obtain $\theta^*$ by solving the following optimization problem:
	\begin{align} \label{eq:polEval}
		\begin{split}
			\theta^* \triangleq \arg\min_\theta \  &\|D(\theta)\|_2 + \alpha \|\theta - \bar{\theta}\|_2, \\
			\text{s.t.} \quad &Q_\phi^\theta \geq 0
		\end{split}
	\end{align}
	where $D(\theta) \triangleq [d_0(\theta), \dots, d_{\timeHorzRL-1}(\theta)]$. The term $\|\theta - \bar{\theta}\|_2$ is a regularizer and $\alpha$ is a gain. The values of $\bar{\theta}$ and $\alpha$ are specified in step 3) of Algorithm~\ref{alg:dataDrivePolIter}. The non-negativity constraint on the approximate Q-function is imposed since the Q-function is a discounted sum of non-negative terms~\eqref{eq:QfuncFixPol}. How it is enforced is described in Section~\ref{sec:ApproxiArchtec}. The solution to~\eqref{eq:polEval} results in $Q^{\theta^*}_\phi$, which is an approximation to $Q_\phi$. The quantity $Q^{\theta^*}_\phi$  can be used to obtain an improved policy, denoted $\phi^+$, through
	\begin{align} \label{eq:polUpdate}
		\phi^+(x) = \arg\min_{u\in\inputSet(x)}\ Q^{\theta^*}_\phi(x,u), \quad \text{for all} \quad x \in \stateSet.
	\end{align}
	This process of policy evaluation~\eqref{eq:polEval} and policy improvement~\eqref{eq:polUpdate} are repeated. This iterative procedure is described formally in Algorithm~\ref{alg:dataDrivePolIter}, with $\numPolImp$ denoting the number of policy improvements.
	\begin{algorithm} 
		\SetAlgoLined
		\KwResult{An approximate optimal policy $\phi^{\numPolImp}(x)$. }
		\KwIn{$\timeHorzRL$, $\theta^0$, $\numPolImp$, $\beta > 1$}

		\For{$j=0,\dots, \numPolImp-1$ }{
			\medskip
			
			\textbf{1)} Obain input sequence $\{u^j_k\}_{k=0}^{\timeHorzRL-1}$, initial state $x^j_0$, and state sequence $\{x^j_k\}_{k=1}^{\timeHorzRL}$. \ 
			
			\medskip
			
			\textbf{2)} For $k=1,\dots, \timeHorzRL$, obtain: $\phi^{j}(x_k) = \arg\min_{u\in\inputSet(x_k^j)} Q_\phi^{\theta^j} (x^j_k,u)$. \
			
			\medskip
			
			\textbf{3)} Set $\bar{\theta} = \theta^j$ and $\alpha = \frac{j}{\beta}$ appearing in~\eqref{eq:polEval}. 
			
			\textbf{4)} Use the samples $\{u^j_k\}_{k=0}^{\timeHorzRL-1}$, $\{x^j_k\}_{k=0}^{\timeHorzRL}$, and $\{\phi^{j}(x_k)\}_{k=1}^{\timeHorzRL}$ to construct and solve~\eqref{eq:polEval} for $\theta^*$.
			
			\medskip
			
			\textbf{5)} Set $\theta^{j+1} = \theta^*$.
			
			\medskip
		}
		\caption{Data Driven Policy Iteration: Batch mode and off-policy}
		\label{alg:dataDrivePolIter}
	\end{algorithm}	
	
	This algorithm is inspired by: (i) the Batch Convex-Q learning algorithm found in~\cite[Section III]{ConvexLuACC:2021} and (ii) the least squares policy evaluation (LSPI) algorithm~\cite{LeastSquaresLagoudakisJMLR:2003}. The approach here is simpler than the batch optimization problem that underlies the algorithm in~\cite[section III]{ConvexLuACC:2021}, which has an objective function that itself contains an optimization problem. In comparison to~\cite{LeastSquaresLagoudakisJMLR:2003} we include a regularization term that is inspired by proximal methods in optimization that aids convergence, and a constraint to ensure the learned Q-function is non-negative.  
	
	\subsection{Proposed RL controller for \plant} \label{sec:RLarch}
	We now specify the ingredients required to apply Algorithm~\ref{alg:dataDrivePolIter} to obtain a RL controller (i.e., a state feedback policy) for the \plant\ from simulation data. Namely, (1) the state description, (2) the cost function design, (3) the approximation architecture, and (4) the exploration strategy. Parts (1), (2), and (3) refer to the setup of the optimal control problem the RL algorithm is attempting to approximately solve. Part (4) refers to the selection of how the state/input space is explored (step 1 in Algorithm~\ref{alg:dataDrivePolIter}).
	
	\subsubsection{State space description}
	In RL, the construction of the state space is an important feature, and the state is not necessarily the same as the plant state. To define the state space for RL, we first denote $w_k$ as the vector of exogenous variables:
	\begin{align}
		w_k = [(w_k^p)^T, \rho_k,\bar{\rho}_k] \in \R^4.
	\end{align}
	where $\bar{\rho}_k = \frac{1}{\tau}\sum_{t = k - \tau}^{k}\rho_t$ is a backwards moving average of the electricity price. The expanded state for RL is:
	\begin{align}\label{eq:x-def-RL}
		x_k \triangleq [x_k^p, w_k]^T \in \stateSet \triangleq \subset \R^{12}. 
	\end{align}
	Note that with the state defined by~\eqref{eq:x-def-RL}, \emph{ a state feedback policy is implementable} since all entries of $x_k$ can be measured with commercially available sensors (e.g., outside wet-bulb temperature, $T_{oawb}$), or estimated from measurements (e.g., the thermal load from buildings, $\qDotLref$), or known via real-time communication (e.g., the electricity prices, $\rho_k$ and $\bar{\rho}_k$). 
	
	\subsubsection{Design of stage cost}
	The design of the stage cost is also an important aspect of RL. We wish to obtain a policy that tracks the load  $\qDotLref_k$ whilst spending minimal amount of money, as described in section~\ref{sec:contGoal}. Therefore we choose:
	\begin{align}
		\label{eq:5}
		c(x_k,u_k) \triangleq c_k^{\text{E}} + \kappa\left(\qDotL_k - \qDotLref_k\right)^2,
	\end{align}
	where $\kappa$ is a design parameter; $\kappa \gg 1$ will prefer load tracking over energy cost.
	
	\subsubsection{Approximation architecture} \label{sec:ApproxiArchtec}
	We choose the following linear-in-the-parameter approximation of the $Q$ function:
	\begin{align} \label{eq:QfuncApprox}
		Q^{\theta}_{\phi}(x,u) = \sum_{\ell=1}^d\psi_\ell(x,u)\theta_\ell,
	\end{align}
	where $\psi_\ell(x,u)$ are nonlinear basis functions and $\theta \in \R^d$ is the parameter vector. We elect a quadratic basis, so that each $\psi_\ell(x,u)$ is of the form $xu$, $x^2$, or $u^2$. A subset of all possible combinations are included in the basis. More on this subset is provided in Section~\ref{sec:under-the-hood}. We can equivalently express the approximation~\eqref{eq:QfuncApprox} as:
	\begin{align} \label{eq:QfuncMatForm}
		Q^{\theta}_{\phi}(x,u) = [x,u]P_\theta [x,u]^T,
	\end{align}
	for appropriately chosen $P_\theta$. In this form it is straightforward to enforce the constraint in~\eqref{eq:polEval} by enforcing the convex constraint $P_\theta \geq 0$. 
	
	\subsubsection{Exploration strategy} \label{sec:simEnv}
	Exploration refers to how the state/input sequences appearing in step 1) of Algorithm~\ref{alg:dataDrivePolIter} are simulated. We utilize a modified $\epsilon-$greedy exploration scheme. At time step $k$ of iteration $j$, we obtain the input $u_k^j$ from one of three methods: (i) by using the policy in step 2) of Algorithm~\ref{alg:dataDrivePolIter}, (ii) electing uniformly random feasible inputs, and (iii) using a rule-based baseline controller (described in Section~\ref{sec:ruleBased}). The states are obtained sequentially through simulation, starting from state $x^j_0$ for each $j$. The choice to use either of the three controllers is determined by the probability mass function $\nu^j_{\text{exp}} \in \R^3$, which depends on the iteration index of the policy iteration loop: 	
	\begin{align}
		\nu^j_{\text{exp}} = \begin{cases}
			[0,0.1,0.9] & \text{for} \ j\leq 5. \\
			[0.5,0.25,0.25] & \text{for} \ j> 5.
		\end{cases}
	\end{align}
	The entries correspond to the probability of using the corresponding control strategy, which appear in the (i)-(iii) order as just introduced. The rational for this choice is that the BL controller provides ``reasonable'' state input examples for the RL algorithm in the early learning iterations so to steer the parameter values in the correct direction. After this early learning phase, weight is shifted towards the current working policy so to force the learning algorithm to update the parameter vector in response to its actions.
	
	The policy evaluation problem~\eqref{eq:polEval} during training is solved using CVX~\cite{cvx}. The simulation model~\eqref{eq:plant-model-f} to generate state updates, which requires solving a non-convex NLP, is run by  using CasADi and IPOPT~\cite{Andersson2019,wacbie:2006}.
	
	The parameters used for RL training are $\gamma = 0.97$, $d = 36$, $\kappa = 500$, $\beta = 100$, $\timeHorzRL = 432$  and $\numPolImp = 50$. The parameter $\tau$ for the backward moving average filter on the electricity price is chosen to represent $4$ hours. The choice of the 36 basis functions are a bit involved; they are discussed in Section~\ref{sec:under-the-hood}. Because a simulation time step, $k$ to $k+1$, correspond to a time interval of $10$ minutes, $\timeHorzRL = 432$ corresponds to 3 days. The controller was trained with weather and load data for the three days Oct. 10-12, 2011, from the Singapore UWC campus dataset described in Section~\ref{sec:modelCal}. The electricity price data used for training was taken as a scaled version of the locational marginal price from PJM~\cite{PJM:ElecPriceUrl} for the three days Aug. 30 - Sept. 1, 2021.
	
	\subsection{Real time implementation}
	Once the RL controller is trained, it computes the control command $u_k$ in real-time as:
	\begin{align} \label{eq:uk-argMinQ}
		u_k := 	\phi^*(x_k) = \arg\min_{u\in\inputSet(x_k)} \ Q_\phi^{\hat{\theta}}(x_k,u),
	\end{align}
	where $\hat{\theta}$ is the parameter vector learned in Algorithm~\ref{alg:dataDrivePolIter}. This $\hat{\theta}$ needs not to be $\theta^{\numPolImp}$ but the one with the best closed-loop performance, which is explained later in Section~\ref{sec:ResultAndDiscus}. 
	
	Due to non-convexity of the set $\inputSet(x_k)$ and integer nature of $\nCh_k$, the problem~\eqref{eq:uk-argMinQ} is non-convex and integer-valued. We solve it as follows: for each possible value of $\nCh_k$, we solve the corresponding continuous variable non-linear program using CasADi/IPOPT~\cite{Andersson2019,wacbie:2006}, and then choose the minimum out of ($\nChMax$+1) solutions by direct search. Direct search is feasible because of of $\nChMax$ for \plant s is a small number in practice ($\nChMax=7$ in our simulated example).

	\section{Proposed Model Predictive Controller}\label{sec:MPC}
	Recall that a straightforward translation of \eqref{eq:ControlGoal_infinite} to MPC will require solving the following problem at every time index $k$ (here we only describe the one at $k=0$ to avoid cumbersome notation):
	\begin{align} \label{eq:ControlGoal_finite}
		\min_{\{u_k\}_{k=0}^{\timeHorzMPC-1}} \ &\sum_{k=0}^{\timeHorzMPC-1}c_k^{\text{E}}, \\
		\text{s.t.} \quad &x^p_{k+1} = f(x^p_k,u_k,w^p_k), \ x^p_0 = x, \nonumber\\
		& x^p_k \in \stateSet^p(w^p_k), \quad u_k \in \inputSet(x^p_k,w_k) \nonumber \\
		& \qDotL_k(x^p_k,u_k) = \qDotLref_k, \nonumber
	\end{align}	
	where $c_k^{\text{E}}$ is defined in \eqref{eq:ckE}, and $\timeHorzMPC$ is the planning horizon. Even for a moderate planning horizon $\timeHorzMPC$ the optimization problem~\eqref{eq:ControlGoal_finite} will be a large MINLP. We now describe an algorithm that uses a dynamic model of the \plant\ to approximately solve \eqref{eq:ControlGoal_finite} without needing to solve an MINLP or even an MILP. This algorithm, which we call \emph{MBOC}, for \emph{Model Based (sub) Optimal Controller}, is then used to implement MPC by repeatedly applying it in a receding horizon setting as new forecasts of external disturbances become available.
	
	The first challenge we have to overcome is not related to the mixed-integer nature of the problem but is related to the complex nature of the dynamics. Recall from Section~\ref{sec:plantModel} that the dynamic model, i.e., the function $f $ in the equality constraint $x_{k+1} = f(\cdot)$ in~\eqref{eq:plant-model-f} is not available in explicit form; rather the state is propagated in the simulation by solving an optimization problem. Without an explicit form for the function $f(\cdot)$, modern software tools that reduce the drudgery in nonlinear programming, namely numerical solvers with automatic differentiation, cannot be used.
	
	We address this challenge by substituting the implicit equality constraint $x^p_{k+1} = f(x^p_k,u_k,w^p_k)$ in~\eqref{eq:ControlGoal_finite} with the underlying constraints $\Omega_k(\cdot)$ in~\eqref{eq:plantDyn-mainbody}, and add the objective of~\eqref{eq:plantDyn-mainbody} to the objective of~\eqref{eq:ControlGoal_finite}. The modified problem becomes:
	
	\begin{align}\label{eq:MPCfinal}
		\min_{u_k} & \sum_{k=0}^{\timeHorzMPC-1} c_{k}^{\text{E}} + r_1\|\qDotL_k-\qDotLref_k\|^2 + r_2\|\TChWS_{k+1}-\TChWS_{\set}\|^2 \\
		&  + r_3\|\TCWS_{k+1}-\TCWS_{\set}\|^2, \nonumber \\
		\text{s.t.} \quad & x^p_{k+1} \in \Omega_k(x^p_k,u_k,w^p_k), \ x^p_0 = x, \nonumber\\
		& x^p_k \in \stateSet^p(w^p_k), \quad u_k \in \inputSet(x^p_k,w_k). \nonumber
	\end{align}
	Since the input $n_{k}^{\ch}$ takes integer value in the set $\{0,1,\ldots,\nChMax\}$, the problem \eqref{eq:MPCfinal} is still a high-dimensional MINLP.
	
	The proposed algorithm to approximately solve \eqref{eq:MPCfinal} without using an MINLP solver or an MILP relaxation consist of three steps. These are listed below in brief, with more details provided subsequently.
	\begin{enumerate}
		\item The integer variable $n^{\ch} \in [0,1,\ldots,\nChMax]$ is relaxed to a continuous one $n^{\ch,c}\in [0,\nChMax]$. The relaxed problem, an NLP, is solved using an NLP solver to obtain a locally optimal solution. In this paper we use IPOPT (through CasADi) to solve this relaxed NLP.
		\item The continuous solution $\{n^{\ch,c}_k\}_{k=0}^{\timeHorzMPC-1} \in \R^{\timeHorzMPC}$, resulting	from Step 1, is processed by using Algorithms \ref{alg:movingAverageRound}
		and \ref{alg:switchingConstriant} to produce a transformed solution that is integer-valued, which is denoted by $\{n^{\ch,d}_k\}_{k=0}^{\timeHorzMPC-1}$. 
		\item In Problem \eqref{eq:MPCfinal}, the input $\{n_{k}^{\ch}\}_{k=0}^{\timeHorzMPC-1} = \{n_{k}^{\ch,d}\}_{k=0}^{\timeHorzMPC-1}$ is fixed at the values obtained in Step 2, and the resulting NLP is solved again. The resulting solution is called the model-based sub-optimal solution (MBOC).
	\end{enumerate}
	
	In the sequel, we will refer to a vector with non-negative integer components, $x \in \Z^n$, as an n-length \emph{discrete signal}. For a discrete signal $x \in \Z^n$, the number of switches, $N_{\switch}$, is defined as the number of times two consecutive entries differ: $N_{\switch}:= \sum_{i=1}^{n-1} I(x_i - x_{i+1})$, where $I(\cdot)$ is the indicator function: $I(0)=0$ and $I(y)=1$ for $y\neq 0$.
	
	The continuous relaxation in Step 1 is inspired by branch and bound algorithms for solving MINLPs, since such a relaxation is the first step in branch and bound algorithms. However, a simple round-off based method to convert the continuous variable $n^{\ch,c}$ to a discrete one leads to a high number of oscillations in the solution. This corresponds to frequent turning on and off of one or more chillers, which is detrimental to them. 
	
	Step 2 converts the continuous solution from Step 1 to a discrete signal, and involves multiple steps in itself. The first step is Algorithm \ref{alg:movingAverageRound}, which filters the signal $n^{\ch,c}$ with a modified moving average filter with a two hour window  (corresponding to 12 samples with a 10 minute sampling	period) and then rounding up the filtered value to the nearest integer. Thus by operating the moving average filter on $n^{\ch,c}$ one obtains	a discrete signal for the chiller command $n^{\ch,f}=\text{moving\_average\_round(\ensuremath{n^{\ch,c}})}$.

	\begin{algorithm}[h]
		\textbf{Input}: Signal ${\bf x}\in\Z^{n}$, $w \in \Z^+$
		(window length)
		
		
		\begin{itemize}
			\item for i=1:$w$
			
			$\qquad$${\bf x}_{d}[i]$ = $\lceil \text{mean}({\mathbf x}[1:i+w/2]) \rceil$
			
			end
			\item for $i=w/2+1:n-w/2$
			
			$\qquad{\bf x}_{d}[i]$ = $\lceil \text{mean} ({\bf x}[i-w/2:i+w/2]) \rceil$
			
			end
			\item for $i=n-w/2+1:n$
			
			$\qquad{\bf x}_{d}[i]$ = $\lceil \text{mean} ( {\bf x}[i-w/2:end]) \rceil$
			
			end
		\end{itemize}
		
		\textbf{Output}: Discrete signal ${\bf x}_{d}$
		
		\caption{\label{alg:movingAverageRound} ${\bf x_{d}}=\text{moving\_average\_round}({\bf x})$}
	\end{algorithm}
	
	The rounding moving average filter typically does not reduce the switching frequency sufficiently. This is why an additional step, Algorithm \ref{alg:switchingConstriant}, described below, is used to operates on this signal and produce the output $n^{\ch,d}:=\text{reduce\_switching(\ensuremath{n^{\ch,f}})}$ that has fewer switches. 
	
	\begin{algorithm}[h]
		\caption{\label{alg:switchingConstriant} $ \mathbf{x}_{rs}=\text{reduce\_switching}( \mathbf{x} )$}
		\textbf{Input}: Discrete signal ${\bf x}\in\Z^{n}$ and $w \in
		\Z^+$ (window length)
		
		\textbf{ 1:} Obtain indices of the entries
		of $\mathbf{x}$ that are not to be changed, ${\bf
			index\_freezed}$, as follows:
		\begin{itemize}
			\item Initialize ${\bf index\_freezed}$ = zeros(n,1) \# Array of dimension n with all
			entries zero
			\item for $i=1:n$
			
			$\qquad$if $N_{\switch}(\mathbf{x}[i-w:i])$ = 0
			
			$\qquad\qquad$${\bf index\_freezed}[i] \leftarrow 1$
			
			$\qquad$end
		\end{itemize}
		
		\textbf{ 2:}  Initialize $\mathbf{x}_{rs}$: $\mathbf{x}_{rs}[i] = \mathbf{x}[i]$ for each $i$ such that {\textbf{index\_freezed}[i]=1}.
		
		\textbf{ 3:}
		For each $i$ in ${\bf index\_freezed}$ which is 0:
		\begin{itemize}
			\item Find all the consecutive 0 entries                                
			till the next 1. Let these indices be called $I_{s}^{i}$, and define $y_i = \lceil mean({\bf x}[I_{s}^{i}]) \rceil$.
			\item Set $\mathbf{x}_{rs}[j] \leftarrow y_i$ for every $i \in I_{s}^{i}$.
			\item Set ${\bf index\_freezed}[I_{s}^{i}] \leftarrow 1$
		\end{itemize}
		end
		
		\textbf{Output}: $\mathbf{x}_{rs}$
	\end{algorithm}

	The need for Step 3 is that the chiller command $\{n^{\ch,d}\}$ at the end of the second step, together with other variables in the solution vector from Step 1, may violate some constraints	of the optimization problem \eqref{eq:MPCfinal}. Even if $\{x^p_{k+1}\}$ and $\{n^{\ch,d}\}$ are feasible, the resulting control commands may not track the cooling load adequately. Step 3 ensures a feasible solution and improves tracking.
	
	\emph{Forecasts:} Implementation requires availability of the forecasts of disturbance $w_k^p$, i.e., cooling load reference and electricity price, over the next planning horizon. There is a large literature on estimating and/or forecasting loads for buildings and for real-time electricity prices; see~\cite{bracha:2002,GuoAggregationEnB:2020,OldewurtelCDC:2010} and references therein. The forecast of $\TOAWB_k$ is available from National Weather Service~\cite{NWS}. We therefore assume the forecasts of the three disturbance signals, $\qDotLref_k$, $\TOAWB_k$ and $\rho_k$, are available to the MPC controller at each $k$.
	
	\section{Rule-based Baseline Controller}\label{sec:ruleBased}
\begin{figure}[h]
		\centering    
		\includegraphics[width=1.1\linewidth, height=1.5\linewidth]{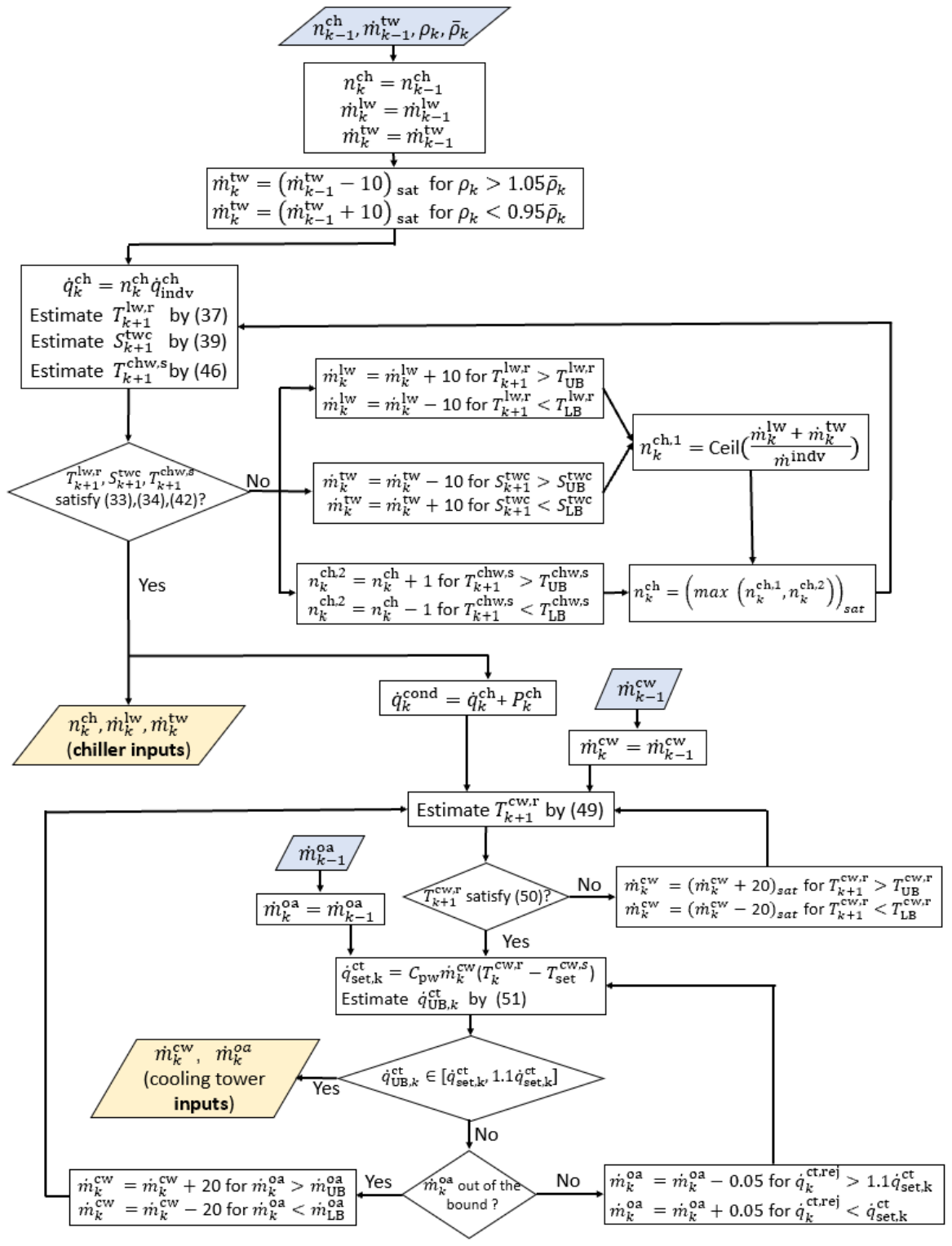}
		\caption{Baseline Controller}. 
		\label{fig:BLController}
	\end{figure}

	In order to evaluate the performances of the RL and MPC controller, we will compare them to a rule-based baseline controller (BL). The proposed baseline controller is designed to utilize the TES and time varying electricity prices (to the extent possible with heuristics) to reduce energy costs. The RL controller and baseline controller have the same information about the price: the current price $\rho_k$ and a backward moving average $\bar{\rho}_k$. At each timestep $k$, the baseline controller determines the control command $u_k=[\mDotLW_k, \mDotTW_k, \nCh_k, \mDotCW_k, \mDotOA_k]^T$ following the procedure shown in Figure~\ref{fig:BLController}. The flowcharts are explained in Section~\ref{sec:BL_CH} and~\ref{sec:BL_CT}. The subscript ``sat'' indicates the variable is saturated at its upper or lower bound; the numerical values of the bounds used in simulations are shown in Table~\ref{tab:simParams}. 
	
	\subsection{For chilled water loop}\label{sec:BL_CH}
	\begin{enumerate}
		\item At time step $k$, $\nCh_{k}$, $\mDotLW_k$ and $\mDotTW_k$ are initialized to  $\nCh_{k-1}$, $\mDotLW_{k-1}$ and $\mDotTW_{k-1}$.
		\item The BL controller increases or decreases $\mDotTW_k$ by a fixed amount (10 kg/sec.) if $\rho_k$ is 5\% lower or higher than $\bar{\rho}_k$ in order to take the advantage of time-varying electricity price. 
		\item The BL controller estimates $\TLWR_{k+1}$,  $\TChWS_{k+1}$, $S^{\twc}_{k+1}$ by~\eqref{eq:T_lwr}, by~\eqref{eq:CHHeatBalance}, and \eqref{eq:S_TW} under the assumption of $\mDotBP = 0$ and $\qDotCh_{k} = \nCh_{k}\qChCapIndiv$. 	If $\TLWR_{k+1}$,  $\TChWS_{k+1}$, $S^{\twc}_{k+1}$ are within their bounds, the already computed control command for the chilled water loop is executed. Otherwise, the controller repeatedly increases/decreases $\mDotLW_k$ and $\mDotTW_k$ by a fixed amount (10 kg/sec), and $\nCh_k$ by 1 until $\TLWR_{k+1}$, $\TChWS_{k+1}$, and $S^{\twc}_{k+1}$ are within their bounds. Since $\mDotLW_k + \mDotTW_k$ determines the minimum required $\nCh_k$ through~\eqref{eq:T_sw} and~\eqref{eq:mDot_BP}, the final $\nCh_k$ is readjusted to meet the minimum required $\nCh_k$. 
	\end{enumerate}
	
	\subsection{For cooling water loop}\label{sec:BL_CT}
	\begin{enumerate}
		\item $\mDotCW_k$ and $\mDotOA_k$ are initialized to $\mDotCW_{k-1}$ and $\mDotOA_{k-1}$.
		\item The BL controller estimates $\TCWR_{k+1}$ by~\eqref{eq:T_CWR} by assuming a fixed $eta$. This value is to be estimated from historical data.	If $\TCWR_{k+1}$ is above/below its bound, $\mDotCW_k$ is increased/decreased by a fixed amount (20 kg/sec.) repeatedly until $\TCWR_{k+1}$ is within its bound.
		\item Once $\mDotCW_k$ is determined, the capacity of cooling tower $\qDotCT_{\text{UB},k}$ and the required cooling $\qDotCT_{\text{set},k}$ that cools down $\TCWR_{k}$ to $\TCWS_{\text{set}}$ is computed. If $\qDotCT_{\text{set},k} \leq \qDotCT_{\text{UB},k} \leq 1.1\qDotCT_{\text{set},k}$, then the control command for the cooling water loop computed so far is executed. If $\qDotCT_{\text{UB},k} <\qDotCT_{\text{set},k}$ or $\qDotCT_{\text{UB},k} > 1.1\qDotCT_{\text{set},k}$, $\mDotOA_k$ is increased or decreased by a fixed amount (0.05 kg/sec.). Since $\qDotCT_{\text{UB},k}$ depends on the ambient wet-bulb temperature $\TOAWB_k$ (see equation~\eqref{eq:q_CT_rej} ), there can be a case that $\qDotCT_{\text{UB},k}$ cannot satisfy $\qDotCT_{\text{set},k} \leq \qDotCT_{\text{UB},k} \leq 1.1\qDotCT_{\text{set},k}$ even when $\mDotOA_k$ is already at its bound. In this case $\mDotCW_k$ is varied by a fixed amount (20 ke/sec.) repeatedly until $\TCWR_{k+1}$ and $\qDotCT_{\text{UB},k}$ are within their bounds. 
	\end{enumerate}
	\begin{table}
		\centering
		\caption{Simulation Parameters}
		\setlength{\arrayrulewidth}{0.05cm}
		\begin{tabular}{ l c c | l c c} 
			\hline 
			Parameter & Unit & value & Parameter & Unit & value \\ \hline		
			$t_s$ & minutes & 10 & $\frac{ t_s \timeHorzMPC}{60}$ & hours & 24 \\		
			$\frac{\tau t_s}{60}$ & hours & 4 & $\frac{w t_s}{60}$ & hours  & 2 \\
			$n^{\ch}_\text{max}$ & N/A & 7 & $\mDotTW_{\text{max}}/\mDotTW_{\text{min}}$ & $\frac{\text{kg}}{\text{sec}}$ & 30/-30 \\
			$\mDotLW_\text{max}/\mDotLW_\text{min}$ & $\frac{\text{kg}}{\text{sec}}$ & 350/20 & 	$\mDotCW_{\text{max}}/\mDotCW_{\text{min}}$ & $\frac{\text{kg}}{\text{sec}}$ & 	300/20 \\
			$ {S_{\text{max}}^{\twc/\tww}} $ & N/A & 0.95 & ${S^{\twc/\tww}_{\text{min}}} $ & N/A & 0.05  \\
			\hline
		\end{tabular}
		\label{tab:simParams}
	\end{table}
	
	\section{Performance evaluation}\label{sec:eval}
	\subsection{Simulation setup}	
	Simulations for closed loop control with RL, MPC and baseline controllers are performed for the week of Sept. 6-12, 2021, which we refer to as the \emph{testing week} in the sequel. The weather data for the testing week is obtained from the Singapore data set described in Section~\ref{sec:modelCal}. The real-time electricity price used is a scaled version of PJM's locational margin price for the same week~\cite{PJM:ElecPriceUrl}. Other relevant simulation parameters are located in Table~\ref{tab:simParams}. There is no plant-model mismatch in the MPC simulations. In particular, since the forecasts of disturbance signals are available in practice (see the discussion at the end of Section~\ref{sec:MPC}), in the simulations the MPC controller is provided error-free forecasts in the interest of simplicity.
	
	\emph{We emphasize that the closed loop results with the RL controller presented here are ``out-of-sample'' results, meaning the external disturbance $w_k$ (weather, cooling load, and electricity price) used in the closed loop simulations are different from those used in training the RL controller.}
	
	Four performance metrics are used to compare the three controllers. The first is the energy cost incurred. The second is the Root Mean Square Error (RMSE) in tracking the cooling load reference:
	\begin{align}\label{eq:e-rms-def}
		e_{RMSE} \eqdef \left( \frac{1}{N_{\sim}-1}\sum_{k=1}^{N_{\sim}}(\qDotLref_k - \qDotL_k)^2\right)^\frac{1}{2},
	\end{align}
	where $N_{\sim}$ is the duration for which closed loop simulations are carried out, which in this paper is $1008$ (corresponding to a week: $7\times 24 \times 6$). The third is the number of chiller switches over the simulation period:
	\begin{align}
		n_{\text{switch}}^{\ch} \eqdef \sum_{k=1}^{N_{\sim}-1} |n^{\ch}_{k+1} - n^{\ch}_{k}|.
	\end{align}
	The fourth is control computational time during closed loop simulations.
	
	\subsection{Numerical Results and Discussion} \label{sec:ResultAndDiscus}
	A summary of performance comparisons from the simulations is shown in Table~\ref{tab:perf-comp}. All three controllers meet the cooling load adequately (more on this later), and both the RL and MPC controllers reduce energy cost over the baseline by about the same amount (16.8\% for RL vs. 17.8\% for MPC). These savings are comparable with those reported in the literature for MPC with MILP relaxation and RL. 
	
\begin{figure}[h]
		\centering
		\includegraphics[width=1\linewidth]{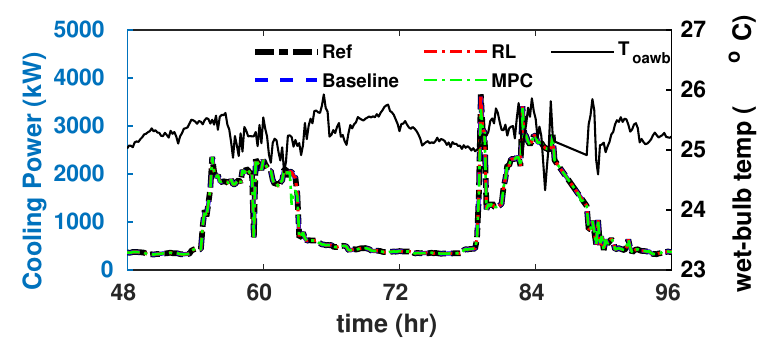}
		\caption{Load tracking performances of the MPC, RL, and BL controllers:  The ``Ref'' is cooling required $\qDotLref_k$.}
		\label{fig:q_L}
	\end{figure}

	In terms of tracking the reference load, both RL and MPC again perform similarly while the baseline controller performs the best in terms of standard deviation of the tracking error; see Figure~\ref{fig:q_L} and Table~\ref{tab:perf-comp}. The worst tracking RMSE is 61 kW, which is a small fraction of the mean load (1313 kW). Thus the tracking performance is considered acceptable for all three controllers. The fact that the baseline performs the best in tracking the cooling load is perhaps not surprising since it is designed primarily to meet the required load and keep chiller switching low, with energy cost a secondary consideration.
	
	In terms of chiller switches, the RL controller performs the worst; see Table~\ref{tab:perf-comp}. This is not surprising because no cost was assigned to higher switching in its design. The MPC performs the best in this metric, again most likely since keeping switching frequency low was an explicit consideration in its design. Ironically, this feature was introduced into the MPC controller after an initial design attempt without it led to extremely high switching frequency. 
	
	In terms of real-time computation cost, the baseline performs the best, which is not surprising since no optimization is involved. The RL controller has two orders of magnitude lower computation cost compared to MPC. The computation time for all controllers is well within the time budget, since control commands are updated every 10 minutes.
	
	\begin{table*}[h!]
		\begin{centering}
			\caption{Comparison of RL, MPC, and baseline controllers (for a week-long simulation).}
			\label{tab:perf-comp}
			\begin{tabular}{|c|c|c|c|c|}
				\hline
				& Total cost (\$) &  $e_{RMSE}$ (kW) & No. of switches & Control computation time (sec, $\mu \pm \sigma$) \tabularnewline 
				\hline 
				Baseline & 3308  & 4.14e-4 & 45  & 8.9e-5 $\pm$ 3.9e-4 \tabularnewline
				\hline 
				RL & 2752  & 1.85 & 114 & 0.32 $\pm$ 0.01   \tabularnewline
				\hline 
				MPC & 2709  & 61.38 & 65 & 27.33 $\pm$ 5.99\tabularnewline
				\hline 
			\end{tabular}
			\par
		\end{centering}
		
	\end{table*}
	
\begin{figure}[h]
		\centering
		\includegraphics[width=1\linewidth]{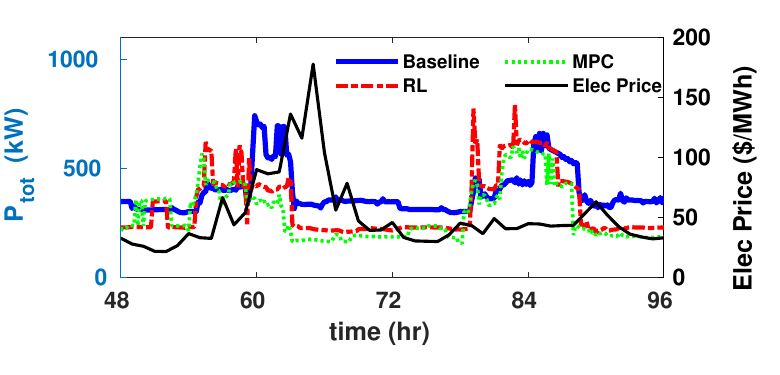}
		\caption{Power consumption vs. real-time electricity price for the MPC, RL, and BL controllers.}
		\label{fig:P_tot}
	\end{figure}

\begin{figure}[h]
		\centering
		\includegraphics[width=1\linewidth]{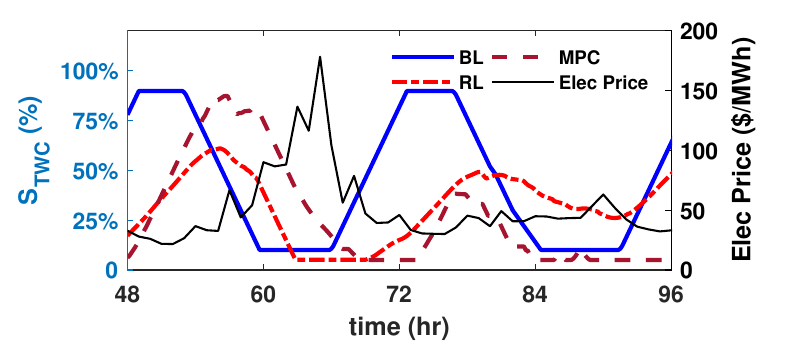}
		\caption{TES cold water volume vs. real-time electricity price for the MPC, RL, and BL controllers. }
		\label{fig:mDot_TW}
	\end{figure}

	{\bf Deeper look:} Simulations are done for a week, but the plots below show only two days to avoid clutter. The cost savings by RL and MPC controller come from their ability to use the TES to shift the peak electric demand to periods of low price better than the baseline controller; see Figure~\ref{fig:P_tot}. The MPC controller has the knowledge of electricity price along whole the planning horizon, and thus achieves the most savings. The cause for cost saving difference between BL and RL controllers is that the RL controller learns the variation in the electricity price well, or at least better than the BL controller. This can be seen in Figure~\ref{fig:mDot_TW}. The RL controller always discharges the TES ($S^{\twc}$ drops) during the peak electricity price while the baseline controller sometimes cannot do so because the volume of cold water is already at its minimum bound. The BL controller discharges the TES as soon as the electricity price rises, which may result in insufficient cold water stored in the TES when the electricity price reaches its maximum. While both the RL and BL controllers are forced to use the same price information (current and a backward moving average), the rule-based logic in the baseline controller cannot use that information as effectively as RL. 
	
	
\begin{figure}[h]
		\centering
		\includegraphics[width=1\linewidth]{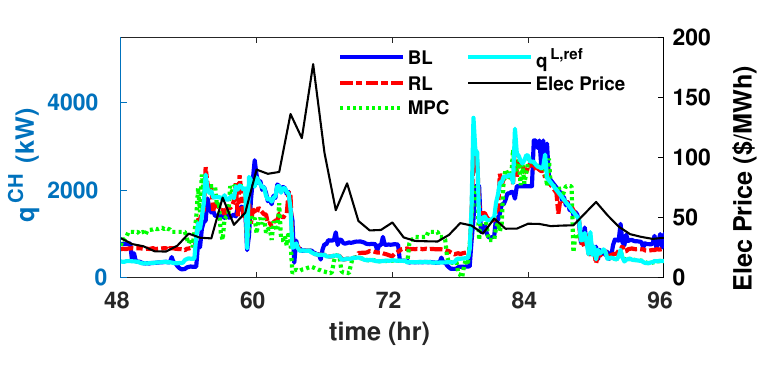}
		\caption{Required cooling load vs. real-time electricity price for the MPC, RL, and BL controllers.}
		\label{fig:q_CH}
	\end{figure}

	An alternate view of this behavior can be obtained by looking at the times when the chillers are turned on and off, since using chillers cost much more electricity than using the TES, which only needs a few pumps. We can see from Figure~\ref{fig:q_CH} that all controllers shift their peak electricity demand to the times when electricity is cheap. But the rule-based logic of the BL controller is not able to line up electric demand with low price as well as the RL and MPC controllers do.
	
\begin{figure}[h]
		\centering
		\includegraphics[width=1\linewidth]{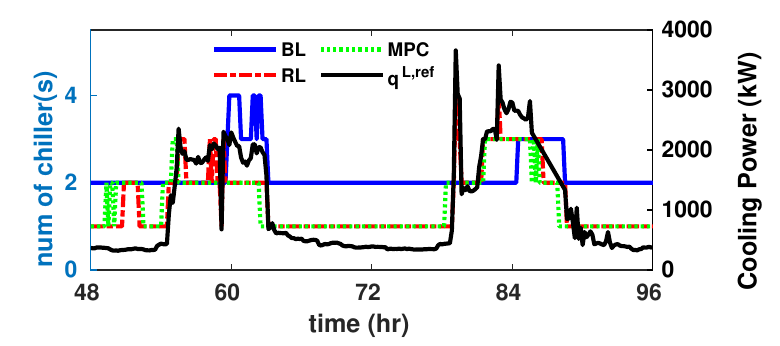}  
		\caption{Number of active chillers vs. real-time electricity price for the MPC, RL, and BL controllers.}
		\label{fig:nChiller}
	\end{figure}

	Another benefit of the RL controller is that it cycles the chillers less than the BL controller even the cost of switching between on-off status of chillers is not incorporated in the cost function; see Figure~\ref{fig:nChiller}. Fast cycling decreases the life expectancy of a chiller greatly.
	
	\section{Under the hood of the RL controller}\label{sec:under-the-hood}
	More insights about why the learned policy works under various conditions can be obtained by taking a closer look at the design choices made for the RL controller. All these choices were the result of considerable trial and error.
	
\begin{figure}
		\centering
		\includegraphics[width=0.72\columnwidth, height=0.62\columnwidth]{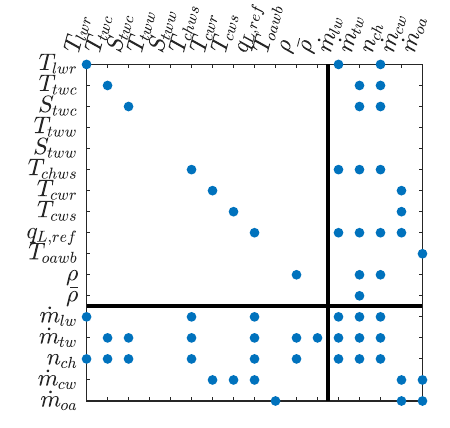}
		\caption{Sparsity pattern of the matrix $P_\theta$ appearing in~\eqref{eq:QfuncMatForm}.}
		\label{fig:sparPattern}
	\end{figure}

	{\bf Choice of basis functions} The choice of basis to approximate the Q-function is essential to the success of the RL controller. It defines the approximate Q-function, and consequently the policy~\eqref{eq:uk-argMinQ}. Redundant basis functions can lead to overfitting, which causes poor out-of sample performance of the policy. We avoid this effect by selecting a reduced quadratic basis, which are the 36 unique non-zero entries shown in Figure~\ref{fig:sparPattern}. Another advantage of reducing the number of basis functions is that it reduces the number of parameters to learn, as training effort increases dramatically with the number of parameters to learn.
	
	The choices for the basis were based on physical intuition about the \plant.  First, basis functions can be simplified by dropping redundant states. One example is $S^{\tww}$. Since $S^{\twc}$ and $S^{\tww}$ are dual terms: $S^{\twc} + S^{\tww} =1$, so one of them can be dropped. Considering that the $S^{\twc}$ reflects the amount of cooling saved in the TES,  we dropped $S^{\tww}$. Another example is the term $T^{\tww}$, which is dropped since it is bounded by $\TLWR$ which is already included in the basis function. Second, if two terms have a strong causal or dependent relationship, e.g., \mDotLW\ and \TLWR\ (see \eqref{eq:T_lwr}), then the corresponding quadratic term \mDotLW\TLWR\ should be selected as an element of the basis. Third, if two terms have minimal causal or dependent relationship, e.g., \mDotOA\ and \mDotTW\ (they are from different equipment and water loops), then the corresponding quadratic term \mDotOA\mDotTW\ should not be selected as an element of the basis.
	
	{\bf Choice of States}  Exogenous disturbances have to be included into the RL states to make the controller work under various cooling load, electricity price, and weather trajectories \emph{that are distinct from what is seen during training}. Without this feature the RL controller will not be applicable in the field.
	
\begin{figure}[h]
		\centering
		\includegraphics[width=0.75\linewidth]{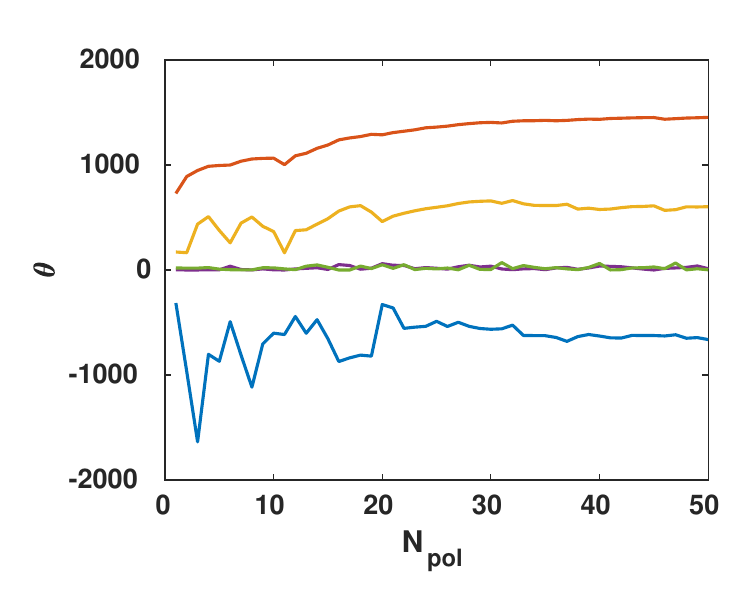}
		\caption{Values of $\theta$ vs. policy iteration index. Only five $\theta$'s are shown to avoid clutter.}
		\label{fig:thetaConverge}
	\end{figure}

	{\bf Convergence of the learning algorithm:} The learning algorithm appears to converge in training, meaning,  $|\theta_k - \theta_{k-1}|$ is seen to reduce as the number of training epochs $k$ increases; see Figure~\ref{fig:thetaConverge}. This convergence should not be confused with convergence to any meaningful optimal policy. The policy learned in $40$th iteration can be a better performing controller than the policy obtained in $50$th iteration. We believe the proximal gradient type method used in learning helps in the parameters not diverging, but due to the same reason it may prevent the parameters from converging to a far away optima. This trade-off is necessary: our initial attempts without the damping proximal term was not successful in learning anything useful. As a result,  after a few policy improvement iterations, every new policy obtained had to be tested by running a closed-loop simulation to assess its performance. The best performing one was selected as  ``the RL controller'', which happened to be the 26th one.
	
	
	{\bf Numerical considerations for training:} Training of the RL controller is an iterative task that required trying many various configurations of the parameters appearing in Table~\ref{tab:simParams}. In particular, we found the following considerations useful.
	\begin{enumerate}
		\item If the value of $\kappa$ is too small, the controller will not learn to track the load $\qDotLref_k$. On the hand, if $\kappa$ is too large the controller will not save energy cost.
		\item The condition number of~\eqref{eq:polEval} significantly affects the performance of Algorithm~\ref{alg:dataDrivePolIter}. However, the relative magnitudes of state and input values are very different, for example, $\qDotL\in[300, 4000]$ (kW) and $S^{TWC}\in[0.05, 0.95]$, which makes the condition number of~\eqref{eq:polEval} extremely large. Therefore, we normalize all magnitudes of state and input values with their average values. With appropriate scaling of the states/inputs, we reduced the magnitude of the condition number from $1\times 10^{20}$ to $1\times 10^3$.
	\end{enumerate}      
	
	\section{Conclusion}\label{sec:conclusion}
	The proposed MPC and RL controllers are able to reduce energy cost significantly, $\approx 17\%$ in a week long simulation, over the rule-based baseline controller. Apart from the dramatically lower real-time computationally cost of the RL controller compared to the MPC, tracking and energy cost performance of the two controllers are similar. This similarity in performance is somewhat surprising. Though both the controllers are designed to be approximations of the same intractable infinite horizon problem, there is nonetheless significant differences between them, especially the information the controllers have access to and the objectives they are designed to minimize. It should be noted that the MPC controller has a crucial advantage over the RL controller in our simulations: the RL controller has to implicitly learn to forecast disturbances while the MPC controller is provided with error-free forecasts. How much will MPC's performance degrade in practice due to inevitable plant-model mismatch is an open question.
	
	Existing work on RL and on MPC tend to lie in their own silos, with comparisons between them for the same application being rare. This paper contributes to such comparisons for a particular application: control of \plant s. Much more remains to be done, such as examination of robustness to uncertainties.
	
	There are several other avenues for future work. One is to explore non-linear bases, such as neural networks, for designing an RL controller. Another is to augment the state space with additional signals, especially with forecasts, which might improve performance. Of course, such augmentation will also increase the cost and complexity of training the policy. Another avenue for improvement in the RL controller is to reduce number of chiller switches. In this paper all the chillers are considered to be the same. An area of improvement is to extend heterogeneous chillers with distinct performance curves, for both RL and MPC. On the MPC front, an MILP relaxation is a direction to pursue in the future.
	
	\bibliographystyle{asmems4}

	\ifshowArxiv
	
	\appendix
	\section*{Appendix: Simulation model of a \plant}\label{sec:sim-model}
	Consider a multi-chiller system with a primary-secondary pumping, a water thermal energy storage (TES) tank, a set of cooling water pumps, and a cooling tower, as is shown in Figure~\ref{fig:ChillerPlant}. We now describe the plant states and their dynamics, along with the outputs. The \plant\ is described in two parts: (i) the chilled water loop and (ii) the cooling water loop. In the sequel, temperature is in Celsius, flowrate is in kg/sec, power is in kW.	
	
	\subsection*{Chilled water loop}\label{sec:sysModel_CHW}
	The inputs to the chilled water loop are: $\mDotLW$, $\mDotTW$, and $\nCh$. Starting from the load, the load water supply is a mixture of chilled waters from both chillers and the TES. The total water mass in the TES is denoted as $M_{\TES}$ (kg). Temporarily using continuous time, and time $t_k$ corresponding to the index $k$, the rate of enthalpy change of load water supply follows: (dropping specific heat of water $C_\pw$ from both sides) 
	\begin{align} \label{eq:firstLawLWS}
		\frac{d}{dt}\Big(M_{\TES}S^{\lw}(t)T^{\lws}(t)\Big)&\Bigg\vert_{t = t_k} \\
		&= \begin{cases} 
			\dot{m}^{\sw}_kT^{\sw}_k - \dot{m}^{\tw}_kT^{\sw}_k, & \dot{m}^\tw_k > 0 \\
			\dot{m}^{\sw}_kT^{\sw}_k - \dot{m}^{\tw}_kT^{\twc}_k, & \dot{m}^\tw_k < 0 
		\end{cases}. \nonumber
	\end{align}
	Using the chain rule to expand the above differential, and supposing $T^{\lws}$ reaches steady state instantly (which yields $\dot{T}^{\lws} = 0$), we have from~\eqref{eq:firstLawLWS} that:
	\begin{align}\label{eq:T_lws}
		T^{\lw,s}_{k} = T^{\sw}_k + \frac{\text{min}(\dot{m}^\tw_k, 0)}{\dot{m}^\lw_k}\Big(T^{\sw}_k - T^{\twc}_k\Big).
	\end{align}
	The load water return temperature $T^{\lw,r}$, by modeling through the heat balance, evolves as:
	\begin{align}
		{T}^{\lwr}_{k+1} &= \frac{1}{C_{\pw}\dot{m}^{\lw}_k}(\qDotL_k + C_{\pw}\dot{m}_k^{\lw}T_k^{\lws}),\label{eq:T_lwr}\\
		0 &\leq \qDotL_k \leq \qDotLref_k, \quad {T}^{\lwr}_k \leq \TLWRMax \label{eq:CCHeatBalance}.
	\end{align}
	The unit delay between the left hand side and the right hand side is to model the non-instantaneous nature of heat exchange between water and air streams, and the transport delay in the chilled water distribution pipes. The bound on ${T}^{\lwr}$ accounts for heat exchanger (the cooling coil in building AHUs) capacity, which varies for difference \plants.
	
	The water thermal energy storage tank is modeled by two sub-tanks: warm water tank and cold water tank. This approach of modeling water TES is natural since a thermocline separates the warm water and cold water in the TES, which is the bedrock for TES to function properly~\cite{ASHRAE_TESguide:2019}. Associated with each tank are two variables: the temperature of the water, $T^\tww_k$ and $T^\twc_k$, and the fraction of water, $S^\tww_k$ and $S^\twc_k$. The input for both tanks is the mass flow rate $\dot{m}_k^\tw$, which represents the water flow rate of two tanks. When $\dot{m}_k^\tw > 0$ this means that at time $k$, $\dot{m}_k^\tw t_s$ (kg) of water is charged into the cold water tank and $\dot{m}_k^\tw t_s$ (kg) has been discharged from warm water tank.  The dynamics for the storage variables are:
	\begin{align} \label{eq:S_TW}
		\begin{cases}
			S^\twc_{k+1} = S^\twc_k + \frac{t_s\dot{m}^\tw_k}{M_{\TES}}  \\
			\\
			S^\tww_{k+1} = S^\tww_k - \frac{t_s\dot{m}^\tw_k}{M_{\TES}}
		\end{cases}
	\end{align}
	with $S^\tww_{k}, S^\twc_k \in [S^\text{min},S^\text{max}] $. The TES is assumed well insulated and no water mixture between two sub-tanks, so that the temperature of the tank is only effected by the temperature of water coming into and out of the tank. For the warm water tank, heat balance reads:
	\begin{align} \nonumber
		M_{\TES}\big(T^\tww_{k+1}S^\tww_{k+1}-T^\tww_{k}S^\tww_{k}\big) = \begin{cases}
			-t_s \dot{m}_k^\tw T^{\lw,r}_k & \dot{m}_k^\tw < 0 \\
			-t_s \dot{m}_k^\tw T^{\tww}_k & \dot{m}_k^\tw > 0
		\end{cases}
	\end{align}
	which can be combined into one equation as:
	\begin{align}\label{eq:T_tww}
		T^\tww_{k+1} = T^\tww_{k} + t_s \frac{\text{min}(\dot{m}^\tw_k,0)}{M_{\TES}S^\tww_k - t_s\dot{m}^\tw_k}\Big(T^\tww_k - T^{\lw,r}_k\Big).
	\end{align}
	The cold water tank derivation is symmetric, and the final result is
	\begin{align}\label{eq:T_twc}
		T^\twc_{k+1} = T^\twc_{k} + t_s \frac{\text{max}(\dot{m}^\tw_k,0)}{M_{\TES}S^\twc_k + t_s\dot{m}^\tw_k}\Big(T^{\sw}_k - T^\twc_k\Big).
	\end{align}
	
	The supply water flow rates are obtained once inputs $\dot{m}^{\lw}_k$ and $\dot{m}^{\tw}_k$ are chosen. The supply water temperature $T^{\sw}_k$, return water temperature $T^{\rw}_k$, and flow rate $\mDotSW$ are then:
	\begin{align}
		T^{\sw}_k &= T^{\chws}_k \label{eq:T_sw}, \quad {\dot{m}}^{\sw}_k = \dot{m}^{\lw}_k + \dot{m}^{\tw}_k, \ \text{and} \\
		T^{\rw}_{k} &= T^{\lw,r}_k + \frac{\text{max}(\dot{m}^\tw_k, 0)}{\dot{m}^\sw_k}\Big(T^\tww_k - T^{\lw,r}_k\Big).	\label{eq:T_rw}
	\end{align}
	
	In a primary-secondary pumping system, the water goes through each active (``on'') chiller is a constant $\dot{m}^{\indv}$. The total chilled water produced, $\dot{m}^{\chw}_k$, may be more than what is required by loads and TES; a one-way bypass valve sends the redundant chilled water, $\dot{m}^{\bp}_k$, from chiller outlet to chiller inlet. Chilled water return $T^{\chwr}_k$ is a mixture of return water and bypass water. The above relationships are summarized as:
	\begin{align}
		\dot{m}^{\chw}_k &= n^{\ch}_k \dot{m}^{\indv}, \ \dot{m}^{\bp}_k = \dot{m}^{\chw}_k - \dot{m}^{\sw}_k, \ \dot{m}^{\bp}_k \geq 0 \label{eq:mDot_BP}, \\
		T^{\chwr}_k &= T^{\rw}_k + \frac{\dot{m}^{\bp}_k}{\dot{m}^{\chw}_k}( T^{\chws}_k-T^{\rw}_k ). \label{eq:T_chwr}
	\end{align}
	Assume all chillers are identical and each chiller has a nominal cooling capacity of $\qChCapIndiv$, then the chilled water supply temperature $T^{\chws}_{k+1}$ coming out of the chiller evaporator is:
	\begin{align}	
		{T}^{\chws}_{k+1} &= T^{\chwr}_{k} -  \frac{\qDotCh_{k}}{C_{\pw}\dot{m}^{\chw}_k},   \label{eq:CHHeatBalance} \\
		0 \leq \qDotCh_{k} &\leq n^{\ch}_k \qChCapIndiv,  \quad   \TChWSMax \geq 	\TChWS_{k+1}  \geq \TChWSMin,  \label{Cons:T_chws}
	\end{align}
	where bound on $\TChWS_{k+1}$ is due to the requirement of proper function of chillers, which varies for difference \plants.
	\subsection*{Cooling water loop} \label{sec:sysModel_CW}
	The inputs to the cooling water loop are: \mDotCW and \mDotOA. Simply speaking, without elaborating on the refrigerant loop in chillers, part (or all) of the heat absorbed by the chilled water at the buildings is transmitted to the cooling water, which reduces the chilled water's temperature from $T_k^{\chwr}$ to $T_k^{\chws}$ and increases cooling water's temperature from  $T_k^{\cws}$ to $T_k^{\cwr}$. The above heat exchange occurs at the chiller condenser. The rate of this heat exchange is denoted by $\qDotCond_k$, with the superscript denoting condenser, and is modeled via heat balance: 
	\begin{align}
		\qDotCond_k = \qDotCh_k + \eta_1P_k^{\ch} \label{eq:qDotCond},
	\end{align}
	where $\eta_1P_k^{\ch}$ is the waste heat from the chiller compressor motors, and $P_k^{\ch}$ is described in~\eqref{eq:P_CH}. Following the heat balance, $\qDotCond_k$ results in an increase of the cooling water:
	\begin{align}
		T_{k+1}^{\cwr} & = \frac{\qDotCond_k}{C_{\pw}\dot{m}_k^{\cw}}+T_k^{\cws}, \label{eq:T_CWR}\\
		T_{k+1}^{\cwr} &\leq \TCWRMax,\label{eq:CondHeatBalance}
	\end{align}
	where bound on ${T}^{\cwr}$ accounts for heat exchanger (the chiller condenser) capacity, which varies for difference \plants.
	
	The cooling capacity of a cooling tower, denoted $\qDotCTUB$, is modeled as~\cite{JinSimplifiedECM:2007}:
	\begin{align}
		\label{eq:q_CT_rej} 
		\qDotCTUBk = \frac{c_1 (\dot{m}_k^{\cw})^{c_3}}{1 +  c_2 \left( \frac{\dot{m}_k^{\cw}}{\dot{m}_k^{\oa}}\right)^{c_3}}(T_k^{\cwr}-T_k^{\oawb}),
	\end{align}
	where $c_1,c_2,c_3$ are empirical constants. The heat rejected through evaporation at the cooling tower is denoted ${\qDotCT}_k$,  and is modeled from the heat exchange at the condenser by a unit delay:
	\begin{align}
		\TCWS_{k+1} & = \TCWR_k - \frac{\qDotCT_k}{C_{\pw}\mDotCW_k },\label{eq:T_CWS}\\
		\TCWS_{k+1}  &\geq \TOAWB_k + 1, \quad	0 \leq \qDotCT_{k} \leq \qDotCTUBk, \label{eq:CTHeatBalance}
	\end{align}
	with a minimum approach 1$^\circ$C.
	
	In order to simulate the dynamics of the \plant, the capacities of all heat exchange devices, specified through the constraints in~\eqref{eq:CCHeatBalance},~\eqref{Cons:T_chws},~\eqref{eq:CondHeatBalance}, and~ \eqref{eq:CTHeatBalance}, need to be satisfied. 
	An efficient way to ensure this is through a constrained optimization. The decision variable $z_k$ for the optimization problem consists of the next system states and some current outputs:
	\begin{align}
		z_k \triangleq \big[(x_{k+1}^p)^T, \ \qDotL_k,\ \qDotCh_k,\ \qDotCT_k\big]^T.
	\end{align}
	This variable $z_k$ is solved for by projecting it onto the dynamics and heat exchange capacities of the system. The system then evolves as:
	\begin{align} \label{eq:plantDyn}
		z^{*}_k = \arg&\min_{z_{k}\in\Omega(x^p_k,w^p_k,u_k)} r_1\|\qDotL_k-\qDotLref_k\| \nonumber\\
		&+ r_2\|\TChWS_{k+1}-\TChWS_{\set}\| + r_3\|\TCWS_{k+1}-\TCWS_{\set}\|,
	\end{align}
	where $\TChWS_{\set}$ and $\TCWS_{\set}$ are user-specified setpoints to reflect system nominal working status; $r_1$, $r_2$, and $r_3$ are to set a trade-off between maintaining system nominal working status and ensuring required cooling is met; set $\Omega(x_k^p,u_k)$ is defined by the dynamics and constraints of the \plant\ system:
	\begin{align} \nonumber
		\Omega(x_k^p,w_k^p, u_k) \triangleq\big\{\mathbf{Z} : \mathbf{Z} \text{ satisfies } \eqref{eq:T_lws}-\eqref{eq:CTHeatBalance}\big\}.
	\end{align} 
	\fi
	
\end{document}